\documentclass[11pt]{article}

\usepackage{arxiv}
\usepackage{soul}

\usepackage[utf8]{inputenc} 
\usepackage[T1]{fontenc}    
\usepackage{url}            
\usepackage{booktabs}       
\usepackage{amsfonts}       
\usepackage{nicefrac}       
\usepackage{microtype}      
\usepackage{lipsum}		
\usepackage{graphicx}
\usepackage{natbib}

\usepackage{changes}      
\definechangesauthor[name = Thorsten Schmidt, color=PineGreen]{TS}
\definechangesauthor[name = Blanka, color=red]{BH}
\definechangesauthor[name = Hans, color=blue]{HB}
\definechangesauthor[name = Yannick, color=cyan]{YL}

\usepackage[inline]{enumitem}
\usepackage{tikz}
\usetikzlibrary{calc}

\usepackage{amsmath, amsfonts, amssymb, amsthm}
\usepackage{thm-restate}
\usepackage{bm}
\usepackage{bbm}
\usepackage{setspace}
\usepackage{graphicx}
\RequirePackage[ruled]{algorithm2e}

\SetKwComment{Comment}{$\triangleright$\ }{}
\makeatletter
    \def\algbackskip{\hskip-\ALG@thistlm}
\makeatother

\SetCommentSty{mycommfont}

\usepackage{caption}
\usepackage{subcaption}
\usepackage{xargs} 

\usepackage{todonotes}
\newcommandx{\yannick}[2][1=]{\todo[linecolor=blue,backgroundcolor=blue!25,bordercolor=blue,#1]{\footnotesize{YL: #2}}}

\newcommand{\Objective}{J}
\newcommand{\Strategy}{a}
\newcommand{\StrategySpace}{\calA}
\newcommand{\ModelSpace}{\mathscr{P}}
\newcommand{\ModelDist}{\frakP}
\newcommand{\ParameterSpace}{\Theta}
\newcommand{\ModelDistSpace}{\ModelSpace_\ParameterSpace}
\newcommand{\UM}{U}

\allowdisplaybreaks

\newcommand{\N}{\mathbb{N}}
\newcommand{\R}{\mathbb{R}}
\newcommand{\E}{\mathbb{E}}
\newcommand{\bbP}{\mathbb{P}}

\newcommand{\bfa}{{a}}
\newcommand{\bfao}{{a}_{\mathrm{oracle}}}
\newcommand{\bfan}{{a}_{\mathrm{plug\text{-}in}}}
\newcommand{\bfar}{{a}_{\mathrm{u-a}}}
\newcommand{\bfarr}{{a}_{\mathrm{u-a}'}}
\newcommand{\bfac}{{a}_{\mathrm{mix}}}
\newcommand{\bfava}{{a}_{\mathrm{var-adj}}}

\newcommand{\pS}{S}

\newcommand{\peps}{\epsilon}

\newcommand{\dest}{\hat{\mu}}
\newcommand{\sest}{\hat{\sigma^2}}

\newcommand{\calF}{\mathcal{F}}

\newcommand{\calA}{\mathcal{A}}
\newcommand{\calN}{\mathcal{N}}

\newcommand{\calU}{\mathcal{U}}

\newcommand{\frakP}{\mathfrak{P}}

\newcommand{\erf}{\mathrm{Erf}}
\newcommand{\erfc}{\mathrm{Erfc}}
\newcommand{\entr}{\mathrm{Entr}}
\newcommand{\mv}{\mathrm{MeanVar}}
\newcommand{\cvar}{\mathrm{CVaR}}

\newcommand{\Timesteps}{\lbrace - N + 1, \ldots, 1 \rbrace}
\newcommand{\timesteps}{\lbrace - N + 1, \ldots, 0 \rbrace}

\DeclareMathOperator*{\sign}{\mathrm{sign}}

\DeclareMathOperator*{\argmax}{\mathrm{arg\,max}}

\makeatletter 

\newcommand*\Neginternal[3]{\mathpalette\Neg@{{#1}{#2}{#3}}}
\newcommand*\Neg@[2]{\Neg@@{#1}#2}
\newcommand*\Neg@@[4]{%
  \mathrel{\ooalign{%
    $\m@th#1#4$\cr
    \hidewidth$\m@th#3{#1}\mkern\muexpr#2*2$\hidewidth\cr
  }}%
}

\newcommand*\negslash[1]{\m@th#1\not\mathrel{\phantom{=}}}
\newcommand*\snegslash[1]{\rotatebox[origin=c]{60}{$\m@th#1-$}}
\makeatother  

\newcommand{\ind}{\mathbbm{1}}
\newcommand{\myind}[1]{\ind_{\lbrace #1 \rbrace}}
\newcommand{\myE}[2]{\E_{#1}\left[ #2 \right]}
\newcommand{\myV}[2]{\mathbb{V}_{#1}\left[ #2 \right]}
\newcommand{\myCond}[3]{\E_{#1}\left[\left. #2 \right\vert #3 \right]}

\newcommand{\mySet}[2]{\left\lbrace #1 : #2 \right\rbrace}
\newcommand{\myset}[2]{{\lbrace #1, \ldots, #2 \rbrace}}

\usepackage{etoolbox}
\newcommand{\addQEDstyle}[2]{\AtBeginEnvironment{#1}{\pushQED{\qed}\renewcommand{\qedsymbol}{#2}}\AtEndEnvironment{#1}{\popQED}}

\newtheorem{thm}{Theorem}[section]

\newtheorem{lem}[thm]{Lemma}

\theoremstyle{remark}
\newtheorem{rem}{Remark}[section]
\addQEDstyle{rem}{$\triangleleft$}
\newtheorem{exa}{Example}[section]
\addQEDstyle{exa}{$\triangleleft$}

\addQEDstyle{example}{$\triangleleft$}
\usepackage{xcolor}

\newcommand{\change}[1]{#1}
\newcommand{\oldchange}[1]{#1}

\usepackage{mathrsfs}

\newcommand{\ccF}{{\mathscr F}}
\newcommand{\ccN}{{\mathscr N}}

\usepackage[hidelinks,colorlinks,urlcolor=black,citecolor=blue,linkcolor=red]{hyperref}
\usepackage{cleveref}       

\crefname{thm}{theorem}{theorems}
\Crefname{thm}{Theorem}{Theorems}
\crefname{ithm}{informal theorem}{informal theorems}
\Crefname{ithm}{Informal Theorem}{Informal Theorems}
\crefname{lem}{lemma}{lemmata}
\Crefname{lem}{Lemma}{Lemmata}
\crefname{exa}{example}{examples}
\Crefname{exa}{Example}{Examples}
\crefname{defn}{definition}{definitions}
\Crefname{defn}{Definition}{Definitions}

\title{Uncertainty-Aware Strategies:\\ A Model-Agnostic Framework for Robust Financial Optimization through Subsampling}
\date{}

\definecolor{PineGreen}{RGB}{1,121,111}
\newcommand{\ccP}{\mathscr{P}}
\usepackage[most]{tcolorbox}
\tcbset{
  commentbox/.style={
    enhanced,
    colback=PineGreen!10,
    colframe=PineGreen,
    coltitle=black,
    boxrule=0.8pt,
    arc=6pt,
    left=6pt,
    right=6pt,
    top=6pt,
    bottom=6pt,
    fonttitle=\bfseries,
  }
}

 \author{{Hans Buehler} \\
 	\small Department of \\\small Mathematics \\
 	\small TU Munich \\
 	\small Munich, Germany \\
 	\And
 	\href{https://orcid.org/0000-0002-6369-7728}{\includegraphics[scale=0.06]{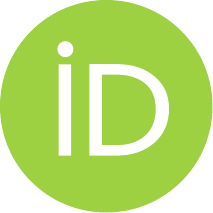}\hspace{1mm}Blanka Horvath} \\
 	\small Mathematical Institute/\\ 
        \small Oxford Man Institute \\
 	\small University of Oxford \\
 	\small Oxford, United Kingdom 
 	\And
 	\href{https://orcid.org/0000-0002-8418-7284}{\includegraphics[scale=0.06]{orcid.pdf}\hspace{1mm}Yannick Limmer}\thanks{Corresponding author, email: \href{mailto:yannick.limmer@maths.ox.ac.uk}{\texttt{yannick.limmer@ox.maths.co.uk}}.} \\
        \small Mathematical Institute/\\ 
        \small Oxford Man Institute \\
 	\small University of Oxford \\
 	\small Oxford, United Kingdom 
 	 	\And
 	\href{https://orcid.org/0000-0002-8418-7284}{\includegraphics[scale=0.06]{orcid.pdf} Thorsten Schmidt} \\ 
 	\small Department of \\
        \small Mathematical Stochastics\\
        \small University of Freiburg  \\
        \small Freiburg, Germany  \\
 }

\renewcommand{\shorttitle}{Uncertainty-Aware Strategies}

\hypersetup{
pdftitle={Uncertainty-Aware Strategies},
pdfsubject={},
pdfauthor={Hans Buehler, Blanka Horvath, Yannick Limmer, Thorsten Schmidt},
pdfkeywords={investing, robust finance, model uncertainty, knightian uncertainty, risk measures},
}

\begin{document}
\maketitle

\begin{abstract}
This paper addresses the challenge of model uncertainty in quantitative finance, where decisions in portfolio allocation, derivative pricing, and risk management rely on estimating stochastic models from limited data. In practice, the unavailability of the true probability measure forces reliance on an empirical approximation, and even small misestimations can lead to significant deviations in decision quality.  
Building on the framework of \cite{KlibanoffMarinacciMukerji05}, we enhance the conventional objective---whether this is expected utility in an investing context or a hedging metric---by superimposing an outer “uncertainty measure”, motivated by traditional monetary risk measures, on the space of models. In scenarios where a natural model distribution is lacking or Bayesian methods are impractical, we propose an ad hoc subsampling strategy, analogous to bootstrapping in statistical finance and related to mini-batch sampling in deep learning, to approximate model uncertainty. To address the quadratic memory demands of naive implementations, we also present an adapted stochastic gradient descent algorithm that enables efficient parallelization. Through analytical, simulated, and empirical studies---\oldchange{including multi-period, real data and high-dimensional examples}---we demonstrate that uncertainty measures outperform traditional mixture of measures strategies and our model-agnostic subsampling-based approach not only enhances robustness against model risk but also achieves performance comparable to more elaborate Bayesian methods.
\end{abstract}


\section{Introduction}

Model uncertainty is a prevalent challenge in quantitative finance. Tasks such as portfolio allocation, derivative pricing, and risk management all hinge on estimating stochastic models for asset prices and risk factors. Whether one employs simpler parametric models or more flexible high-dimensional, machine-learning-based frameworks, the inevitable reality is that the ‘‘true’’ underlying probability measure \(\bbP\) is \oldchange{not known}. Instead, \oldchange{agents typically} rely \oldchange{on alternatives, such as statistical estimation of an estimated measure} \(\hat{\bbP}\). Any \oldchange{estimation error} of \(\hat{\bbP}\) can lead to suboptimal decisions---potentially with severe financial consequences \cite{Knight21,Cont06,GlassermanXu14}.

\begin{exa}
\begin{figure}
    \centering
    \includegraphics[height=5cm]{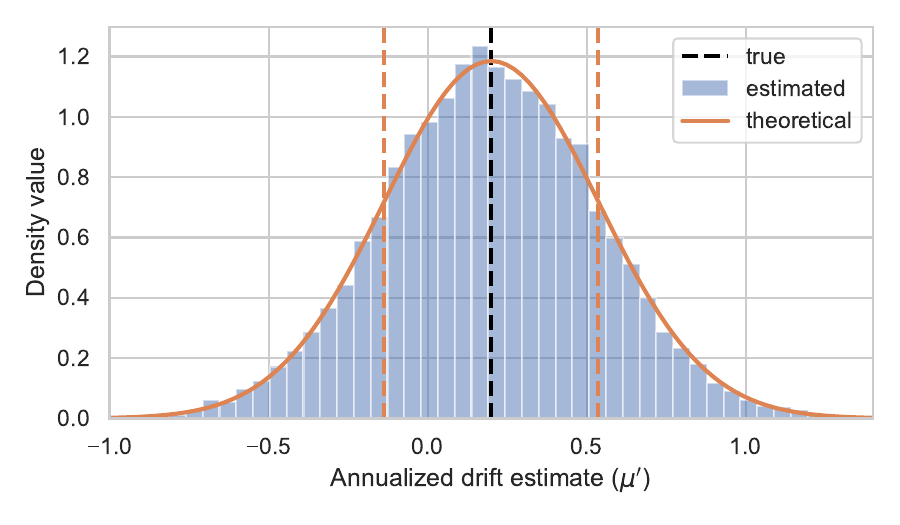}
    \caption{
    Mean estimations based on the observation of $90$ asset returns. 
    The true distribution of the asset returns is Gaussian with mean $20\%$ and volatility of $20\%$.
    The graph shows both the theoretical mean estimation distribution as well as the histogram of estimations based on $10\,000$ paths.
    It is evident that data of $N = 90$ days are not sufficient to obtain accurate estimates of the mean.
    }
    \label{fig:Intro}
\end{figure}
The sensitivity to model misestimation is aptly demonstrated by a simple scenario of estimating the drift of a stock price. Suppose the (annualized) drift \(\mu' = 20\%\) and the volatility \(\sigma' = 20\%\). If daily increments are normally distributed with mean \(\mu' \, dt\) and standard deviation \(\sigma' \sqrt{dt}\), one infers the drift by averaging the historical daily returns:
\[
  \hat{\mu}' \;=\; \frac{1}{N \, dt} \sum_{t=-N+1}^{0} \Delta S_t.
\]
For as few as \(N=90\) daily observations, the standard error can be large enough that \(\hat{\mu}'\) may fail to indicate whether the drift is even positive. Figure~\ref{fig:Intro} (borrowed from a lengthier discussion in the earlier version of this manuscript) illustrates how dramatically \(\hat{\mu}'\) can deviate from \(\mu'\). This situation embodies what is often called \emph{Knightian uncertainty}~\cite{CampbellLoMacKinlay97,Denis06,Cont06,Bielecki18,MuhleKarbe18,TolulopeNeufeldSchmidt19}: the inability to pin down key parameters leads an agent to operate under uncertainty about which ‘‘true’’ model is driving the data.
\end{exa}

Historically, financial models have been calibrated to past market observations by restricting the analysis to a given model class---balancing explainability against mathematical tractability~\cite{Shreve04A, Shreve04B, BlackScholes73, Heston93, HaganKumarLesniewskiWoodward02}. 
More recently, progress in numerical methods and machine learning has shifted the focus towards predictive power, as seen in the development of market generators that fit synthetic data arbitrarily close to observed data~\cite{TavellaRandal00, Duffy10, Gatheral11, Duffy13, HuangChaiCho20, BuehlerGononTeichmannWood19, WieseKnoblichKornKretschmer20, BuehlerHorvathLyonsArribas20, NiSzpruchWieseShujian20, NiSzpruchSabatevidalesWieseXiaoShujian20, BartlDrapeauOblojWiesel21}. 
\oldchange{Up to now}, these powerful methods remain exposed to model uncertainty, as overfitting may result in poor out-of-sample performance~\cite{BaileyBorweinLopezdepradoZhu14}, which underlines the importance of robustification. We focus on methods which are model-free and therefore independent of specific model dynamics.

\paragraph{Our contribution.}
In this paper, we propose an approach to enhance the robustness of optimization tasks in finance by explicitly addressing model uncertainty. \oldchange{Our contribution is trifold:}

\oldchange{Firstly,} our methodology is based on \cite{KlibanoffMarinacciMukerji05} and modifies the standard objective function -- such as a portfolio’s expected utility or a hedging objective -- by adding an \emph{outer} objective function that captures risk over the space of candidate models. We refer to this outer term as an \emph{uncertainty measure}, and show how well-known risk measures (e.g., entropic risk, conditional value-at-risk \cite{FoellmerSchied04,RockafellarUryasev00}) naturally serve this purpose. The resulting objective function thus encompasses:
\begin{itemize}[nosep]
    \item an \emph{inner} component (the original objective) operating under a single model within a chosen class, and
    \item an \emph{outer} component (the uncertainty measure) assessing uncertainty across that class of models.
\end{itemize}

\oldchange{Secondly,} in some settings -- particularly when Bayesian methods are tractable -- one may have a natural probability measure on the model space at hand \cite{HansenSargent01,HansenSargent08}. However, such a measure is not always available or may be prohibitively complex to compute. We therefore develop an ‘‘ad hoc’’ \emph{subsampling} strategy to approximate uncertainty, which is equivalent to the in practice widely used bootstrapping in investment contexts \cite{Efron79} and occurs naturally when sampling mini-batches for deep learning \cite{Goodfellow16,Bottou2010}. The key benefit of this approach is that it is agnostic of the model class used and is available for ``return-'' and ``path-based'' control tasks.

\oldchange{Eventually,} a key computational challenge is that naive implementations of these nested objectives demand quadratic memory in the sample size, thwarting deep learning applications. To overcome this, we propose an adaptation of the stochastic gradient descent (SGD) algorithm that allows to reduce the simultaneous memory burden and admits parallelization \cite{Bottou2012,RobbinsMonro51}. 

\paragraph{Main findings.}
We benchmark our technique against approaches that simply optimize a single mixture measure, combining a distribution over models with their respective laws. 
We show that our method demonstrates improved robustness by flexibly adjusting to model risk; and we show that the mixture measure under canonical choices for the model distribution yields comparatively insufficient robustification. This is best seen in a simple and controlled environment, where analytic solutions are available, but we also demonstrate that our findings extend to simulated and empirical experiments. 

Further, we provide numerical evidence that our suggested model-agnostic \textit{subsampling-based} approach to generate a distribution over the model space, for practical investing and hedging tasks, achieves  performances comparable to more elaborate Bayesian approaches. 
Moreover, we showcase the effect of the $\cvar$-SGD on memory bottle-necks in high-dimensional applications.
Eventually, we demonstrate that when coupled with the deep hedging algorithm \cite{BuehlerGononTeichmannWood19}, and applied to a simulated hedging example, our robustification scheme yields tangible gains.

\paragraph{Related Work.}

\oldchange{
Our work builds upon several strands of research at the intersection of robust optimization, risk-aware decision-making, and data-driven methods in finance.

First, the classical problem of parameter and model uncertainty in portfolio optimization has been extensively studied. Robust optimization techniques~\citep{Bertsimas2011, Goldfarb2003, Kim2018robust} hedge against worst-case scenarios within uncertainty sets. These are practically equivalent to our setting
if the outer uncertainty measure is the woret-case measure. Bayesian approaches~\citep{Black1992, Jorion1986,duembgen2014estimate} incorporate parameter uncertainty probabilistically. Ambiguity-averse models~\citep{Garlappi2007} formalize decision-making under multiple priors. Distributionally robust optimization (DRO) extends these ideas by optimizing decisions under worst-case distributions, using moment uncertainty or Wasserstein balls~\citep{Delage2010, Esfahani2018}. However, these approaches either require restrictive assumptions or complex prior elicitation, limiting their flexibility in data-driven environments.

Our method instead relies on \emph{uncertainty measures}---specifically entropic risk and CVaR~\citep{Artzner1999, RockafellarUryasev00}---to quantify uncertainty in a model-agnostic fashion. This connects to recent work in risk-sensitive reinforcement learning~\citep{Tamar2015, Chow2014, Chow2015}, where coherent risk measures guide decision-making under uncertainty, as well as to the use of CVaR in robust control~\citep{Iyengar2005}. Unlike worst-case optimization, our approach adjusts the degree of robustness continuously via uncertainty aversion parameters. Our approach usually converges against the worst-case for increasing risk aversion.

To approximate model uncertainty without specifying priors, we propose \emph{subsampling} (akin to bootstrapping~\citep{Michaud1989}), drawing parallels to resampling techniques widely used in statistics, decision-making, and deep learning~\citep{Bottou2010, Osband2016}. This allows us to construct empirical distributions over models efficiently---a key methodological innovation relative to fully Bayesian or mixture approaches.

From a computational perspective, our scalable stochastic gradient approach to uncertainty-aware learning draws inspiration from both distributionally robust optimization and recent advances in \emph{deep hedging}~\citep{Buehler2019,buehler2020data,wiese2019deep,murray2022deep,buehler2022deep}, which applies deep learning to complex financial decision problems. 
Like deep hedging, our method addresses high-dimensional decision spaces and leverages machine learning to approximate optimal strategies while remaining robust to model risk.
\oldchange{Model uncertainty of deep hedging has been already addressed in a Knightian setting \cite{LutkebohmertSchmidtSester21} as well as with adversarial methods \cite{Limmer24,Wu23}.}

Finally, our work relates to the growing literature on \emph{risk-sensitive deep learning} in finance and reinforcement learning~\citep{Jiang2017}, where performance metrics explicitly account for tail risks and robustness. Unlike many existing methods, we provide both analytical tractability (in stylized cases) and practical scalability.

In sum, our contribution bridges robust optimization, risk measures, subsampling-based uncertainty quantification, and deep learning---providing a flexible and computationally efficient framework for decision-making under model uncertainty in financial tasks.
}



\section{Uncertainty measures}
In this section, we introduce uncertainty measures and demonstrate how they can be used to address model uncertainty. For this, let $(\Omega,\ccF,\bbP)$ be a probability space and $\StrategySpace$ the space of admissible actions. Suppose we have given an investment opportunity $X: \StrategySpace \times \Omega \to \R$, which for an action $\Strategy \in \StrategySpace$ and the event $\omega \in \Omega$ results in the payoff $X(\Strategy, \omega)$. To align with standard notation in the literature, we omit $\omega$ and refer to $X(\Strategy)$ as a random variable, that is $X(\Strategy) \in L^0(\Omega)$, the space of random variables on $\Omega$. We write $X \sim \bbP$ when we want to express that $X$ follows a distribution implied by the probability measure $\bbP$.

\paragraph{Plug-in strategy.} What we will refer to henceforth as the \emph{plug-in} strategy/policy\footnote{We will use these terms interchangeably, while ``strategy'' is more common in the finance literature, ``policy'' is more often used in the machine learning literature.}
\oldchange{
is choosing the action $\Strategy \in \StrategySpace$ such that the resulting payoff $X(\Strategy)$ optimizes the agent's investment objective $J$ under the assumption that $X(a)$ follows a distribution $\hat \bbP$ which the agent estimates, namely: 
\begin{align}\label{id:Plug-inStrategy}
    \argmax_{\Strategy \in \StrategySpace} \, 
     \Objective(X(\Strategy), \hat \bbP).
\end{align}
}
The word ``model'' in this article generally refers to this problem for a given candidate probability measure $\hat \bbP \in \ModelSpace$. In this sense we use the two synonymously.
The space $\ModelSpace$ is the space of probability measures of $(\Omega,\ccF)$, and, formally, the objective is given as function $J: L^0(\Omega) \times \ModelSpace \to \R$.

\begin{exa}[Investing on normal returns] \label{exa:Plug-inNormal}
Assume the price of a stock is given by $S_0 \in \R$ and an agent invests until time $t = 1$ with the amount $\Strategy \in \R =: \StrategySpace$. The payoff of this contingent claim will be the return $\Delta S_1 := S_1 - S_0$, scaled by the action, i.e. $X(a) := a \Delta S_1$. If the agent's objective is minimizing entropic risk $J :=  \entr_\lambda$  with risk aversion $\lambda \in \R_+$ (see \Cref{def:Entropy}), and if she estimates that $\Delta S_1$ is normal with volatility $\hat\sigma$ and mean $\hat \mu$, then the plug-in strategy is
$
    \argmax_{\Strategy \in \R} 
    \entr_\lambda(a \Delta S_1,\hat \bbP) = {\hat \mu}/{(\lambda \hat \sigma^2)}
$
(see \Cref{lem:SolveEntr}).
\end{exa}
\smallskip

\begin{exa}[Hedging]\label{exa:Hedging}
    Our framework also allows for more complex tasks such as hedging. For a basic version of this, consider a stock price  $S = (S_t)_{t = 0,\dots,T}$ together with 
    an European option represented by its payoff $C$ at time $T$. An admissible action is  described by a predictable process $a = (a_t)_{t = 1,\dots,T-1}$. Here the $a_t$ is the position bought at time $t-1$ and sold at time $t$, hence $a_t$ depends only on information available up to $t-1$, i.e.~is predictable. This results in the final position at time $T$ given by 
    $
       X(a) := \sum_{t=1}^T a_t \Delta S_t - C.
    $
    An agent may now assume that the dynamics of $S$ follow a distribution corresponding to the estimated measure $\hat \bbP$, and is interested in minimizing the variance of the payoff (resulting in the classical delta-hedging), implying the objective
    \begin{align*}
        \argmax_{\Strategy \in \StrategySpace} 
    \left\{
        -
        \mathbb{V}^{\hat \bbP}
        \big[
            \textstyle\sum_{t=1}^T a_t \Delta S_t - C
        \big]
    \right\},
    \end{align*}
    where we denote by $\mathbb{V}^{\hat \bbP}$ the variance under the measure $\hat \bbP$.
\end{exa}

\subsection{The uncertainty-aware strategy.} 
\label{sec:uncertainty aware strategy}
The above strategy is vulnerable to model-misspecification: the agent's trading decision in the example above would change if the parameters of the model were estimated differently or if she did not assume a normal distribution in the first place. 
\oldchange{Moreover, note that there is no functional sensitivity to the uncertainty in the estimate $\hat\bbP$}.

\oldchange{
In the following we consider \change{the probability $\hat \bbP$ itself as a random variable. We therefore introduce a} measurable space $(\ParameterSpace,\ccF_\ParameterSpace)$ \change{and}  denote by $\ccP_\Theta$ the space of probability distributions on $(\Theta,\ccF_\Theta)$. \change{We will use $\theta\in\Theta$ to index 
candidate probability measures $\bbP_\theta$. We use $L^0(\Theta)$ 
to refer} to the space of random variables on $\Theta$.
} In this work, we propose to
\begin{enumerate}[label=(\roman*),nosep]
    \item determine a class of models $(\bbP_\theta)_{\theta \in \ParameterSpace} \oldchange{\subseteq \ModelSpace}$, represented by \change{their} 
    probability measures \change{indexed using} the parameter space $\ParameterSpace$;
    \item \oldchange{consider $\theta$ as random variable in $L^0(\Theta)$} and choose a distribution $\ModelDist \oldchange{ \in \ModelDistSpace}$ on the space of models
    \change{to express their uncertainty with $\theta \sim \ModelDist$};
    \item specify an \emph{uncertainty measure}, which is a mapping $\UM: L^0(\ParameterSpace)\oldchange{\times \ModelDistSpace \to \R}$. 
\end{enumerate}
Accordingly, we propose to extend the strategy from Equation \eqref{id:Plug-inStrategy} to the following 
\textit{uncertainty-aware} strategy
\begin{align}
    \argmax_{\Strategy \in \StrategySpace} 
    \left\{
    \UM\big(\Objective(X(\Strategy), \bbP_\theta), \ModelDist\big) 
    \right\}.\label{id:RobustStrategy}
\end{align}
\begin{exa}[Uncertainty-aware investing in normal returns] \label{exa:RobustNormal} 
\begin{figure}[t]
    \centering
    \hspace{3cm}\includegraphics[height=4.5cm]{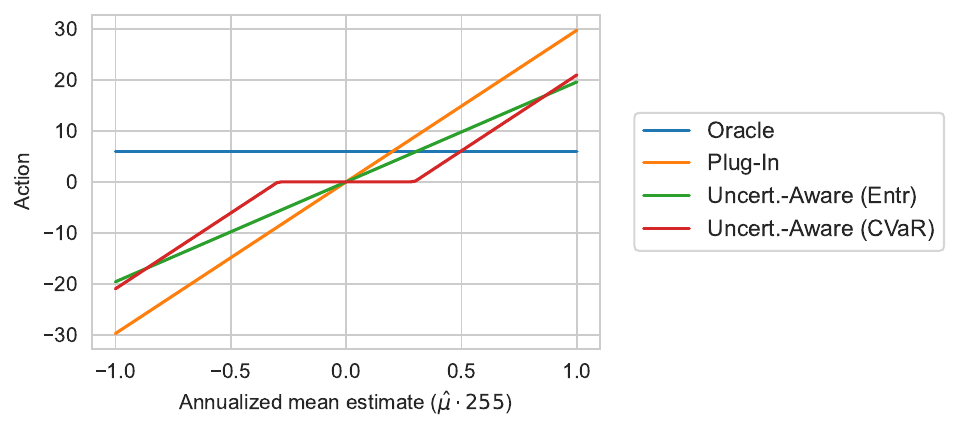}
    \caption{
    Strategies based on the mean estimate of 90 days of observed asset returns (\Cref{exa:Plug-inNormal,exa:RobustNormal}).
    The plug-in strategy is linear in the drift estimate. The uncertainty-aware strategy using $\entr$ as uncertainty measure is linearly as well, however, invests more cautiously---as shown with the lower slope.
    The uncertainty-aware strategy using $\cvar$ as uncertainty measure does not invest for insufficient significant signals, and linearly invests outside that region, parallel to the plug-in strategy.
    The oracle strategy does not depend on the mean estimate, as it has access to the true underlying distribution.
    }
    \label{fig:AnalyticStrategies}
\end{figure}
In the setting of \Cref{exa:Plug-inNormal},  \oldchange{uncertainty in the drift parameter could be incorporated by assuming that $\mu \sim \mathcal{N}(\hat{\mu}, \tau)$ for $\hat \tau \in \R_+$}. With entropy as the uncertainty measure, $U := \entr_{\lambda'}$ 
under uncertainty aversion $\lambda' \in \R_+$ (see \Cref{def:MV})
the uncertainty-aware strategy is given by
\begin{align}
    &\argmax_{\Strategy \in \R} 
    \entr_{\lambda'}\Big(
    \entr_\lambda
    \big(
        a \Delta S_1
        ~ \big| ~
        \Delta S_1 \sim \calN(\mu, \hat \sigma)
    \big)
    ~ \Big| ~
    \mu \sim \calN(\hat{\mu}, \hat \tau)
    \Big),
    \label{id:UAEntrEntr}
\end{align}
see \Cref{lem:OospUAIidEntrEntr} for the analytic representation. (\oldchange{To aid clarity, whenever appropriate, we use the more lengthy notation $\Delta S_1 \sim \ccN(\mu,\sigma)$ to illustrate which random variable is distributed according to which measure.})

Alternatively, one could use $\UM := \cvar_\alpha$ (see \Cref{id:CVaR}) for uncertainty aversion $\alpha \in [0,1]$, \oldchange{where $\alpha=0$ corresponds to the worst-case measure and $\alpha=1$ represents the expectation operator.} This results in the uncertainty aware-strategy 
\begin{align}
    &\argmax_{\Strategy \in \R} 
    \cvar_{\alpha}\Big(
    \entr_\lambda
    \big(
        a \Delta S_1
        ~ \big| ~
        \Delta S_1 \sim \calN(\mu, \hat \sigma)
    \big)
    ~ \Big| ~
    \mu \sim \calN(\hat{\mu}, \hat \tau)
    \Big), 
    \label{id:UACvarEntr}
\end{align}
which is expressed analytically in \Cref{lem:OospUAIidEntrCVaR}. 
Both strategies, together with the plug-in of \Cref{exa:Plug-inNormal} are visualized in \Cref{fig:AnalyticStrategies}.
\end{exa}

The shift from the plug-in strategy to the uncertainty-aware strategy highlights a fundamental change in the agent's task: rather than estimating a single model, the agent now estimates the probability of different models being correct. This raises the question of how to select or approximate these probabilities in a principled manner.

To systematically compare the plug-in and uncertainty-aware strategies from \Cref{exa:Plug-inNormal,exa:RobustNormal}, we first establish a context where canonical choices for such estimations exist. In particular, we explore settings where the estimation of model probabilities follows well-established statistical principles. This provides a concrete foundation for evaluating the advantage of uncertainty-aware strategies in an out-of-sample setting.

Moreover, our analysis will extend to a comparison with an alternative approach: a simple mixture of models. This allows us to highlight key benefits of uncertainty-aware decision-making beyond those achieved by merely averaging over multiple models.

\change{We stress that the uncertainty measures $\entr_{\lambda'}$ and $\cvar_{\alpha}$ are used to control \emph{model uncertainty}. 
When we compare the out-of-sample performance
of different models later on, we will do so under the original objective~$\entr_\lambda$.}


\section{
Comparing strategies in a Gaussian i.i.d.-context}\label{sec:ComparingAnalyticallyIIDNormal}

In this section, we consider a simple \change{Gaussian} i.i.d.\ setting for the task in \Cref{exa:Plug-inNormal,exa:RobustNormal} to analyze their out-of-sample performances analytically. 
This basic setting allows us to demonstrate not only why robustification is beneficial, but also how our approach offers improvements over other approaches.

\paragraph{Setting.} We denote current time by $t=0$ and assume  $N \in \N$ steps of \change{of size $\delta$} of an asset price process $(S_t)_{t =-N,\dots,0}$
have already been observed, which will serve as the basis to estimate a distribution for the next increment $\Delta S_1$.
We assume that increments of the asset price process are i.i.d., hence
\begin{align*}
    \Delta S_{t} \overset{\text{d}}{=} \Delta S_{0} \quad \forall t \in \Timesteps.
\end{align*}
We assume that the true distribution is normal let $\Delta\pS_{t} \sim \calN(\mu, \sigma^2)$ \change{
with drift $\mu\in\R$ and volatility $\sigma>0$ for time steps $\delta$.}\footnote{\change{
We may write $\mu = \delta\mu'$ and $\sigma = \sqrt{\delta} \sigma'$ for annualized drift $\mu'$
and volatility $\sigma'$.}}\change{ Recall that the agent's objective is the entropy with risk-aversion $\lambda\geq 0$.
}

\paragraph{\oldchange{Oracle strategy.}} If the agent knew the true parameters of the distribution of $\Delta S_1$, she would invest exactly $\bfao := {\mu}/{(\lambda \sigma^2)}$, what we refer to as the \emph{oracle} strategy, see \Cref{lem:SolveEntr}. The objective achieved by this strategy is $\entr_\lambda(\bfao \Delta S_1) = {\mu^2}/{(2 \lambda \sigma^2)}$.

\paragraph{Plug-in strategy.} However, 
\oldchange{as $\mu$ and $\sigma$ are not known, they could be estimated by the maximum-likelihood estimators, i.e.\ by}
\begin{align}
	\dest := \frac{1}{N} \sum_{t = -N+1}^0\Delta S_{t},  
    \quad\text{and}\quad
	\sest := \frac{1}{N-1}\sum_{t=-N+1}^0 \Delta S_{t}^2 \label{id:destsest}. 
\end{align}
This results in the agent investing $\bfan := \dest / (\lambda \sest)$ (see \Cref{lem:SolveEntr} with $\nu = \dest, \tau^2 = \sest$). 
Note that if we considered a different sample path, these values would differ. \change{The estimators are
therefore \emph{uncertain} to us.}
As a matter of fact, we know that \change{the standard estimators above have errors}
\begin{align}
 \dest \sim \calN\left(\mu,  \frac{\sigma^2}{N}\right)
    \quad\text{and}\quad
 \sest \sim \frac{\sigma^2}{N-1}\chi^2_{N-1},
 \label{id:esterr}
\end{align}
which means we can evaluate how $\bfan$ performs on $\Delta S_{1} \sim \calN(\mu, \sigma^2)$. To simplify this, assume first $\sest \equiv \sigma^2$.\footnote{Considering $\eqref{id:esterr}$, we can see that the estimation error of $\sest$ is marginal compared to $\dest$.} \Cref{lem:OospPlug-inEmperic} \change{shows that the realized \emph{out-of-sample performance (OOSP)} of the objective function
with the naively estimated parameters is}
\begin{align*}
    \entr_\lambda(\bfan\Delta S_{1}) = \frac{\mu^2}{2\lambda \sigma^2} - \frac{\mu^2 + \sigma^2}{2 N \lambda \sigma^2}.
\end{align*}
This is the objective the agent would achieve if the experiment is repeated infinitely many times using this plug-in strategy.
\change{For sample parameters $\lambda = 0.84$, $N = 140$, $\sigma = 0.2/\sqrt{255}$, and $\mu = 0.2/255$ this yields a negative out-of-sample entropy
performance and is therefore worse than doing nothing.}

\paragraph{Uncertainty-aware strategy.}
In order to obtain the uncertainty aware strategies as outlined in \Cref{exa:RobustNormal}, the agent needs to estimate a distribution of models. A canonical choice is to use the mean estimates \eqref{id:destsest} and use their theoretical estimation distributions \eqref{id:esterr} to deduce a variance. Again, for the sake of analytical solutions, we are only interested in the uncertainty of the drift component.\footnote{Later in numerical experiments we show that all findings hold when considering uncertainty in the variance component as well.} In the notation of \Cref{exa:RobustNormal}, the agent chooses $\hat \mu = \dest$, and $\tau = \sest / N$. With this, we can analytically compute \Cref{id:UAEntrEntr,id:UACvarEntr}.

\begin{restatable}{lem}{OospUAIidEntrEntr}\label{lem:OospUAIidEntrEntr}
    The \change{entropic} uncertainty-aware policy \eqref{id:UAEntrEntr}  has the analytic representation
    \begin{align}
        \bfar =  \frac{\dest}{\lambda  \sest(1 + \frac{\lambda'}{\lambda N})} = \frac{\lambda N}{\lambda N + \lambda'} \bfan, \label{id:AnalyticUAPolicyMV}
    \end{align}
    what results, under the assumption that $\sest \equiv \sigma^2$, in an out-of-sample performance of 
    \begin{align}
       \entr_\lambda(\bfar\Delta S_{1}) \equiv \mv^\bbP_\lambda\left( \bfar \Delta S_{1}\right) 
        &= \frac{x_N}{\lambda \sigma^2} \left(\mu^2 \left( 1 - \frac{x_N}{2} \right) - \frac{x_N(\mu^2 + \sigma^2)}{2N} \right), \label{id:AnalyticUAOOSPMV}
    \end{align}
    where $x_N := \lambda N / (\lambda N + \lambda')$.
\end{restatable}
\begin{proof}
    See \Cref{app:Iid}.
\end{proof}

\begin{restatable}{lem}{OospUAIidEntrCVaR}\label{lem:OospUAIidEntrCVaR}
    The \change{CVaR} uncertainty-aware policy \eqref{id:UACvarEntr} has the analytic representation
    \begin{align*}
        \bfarr = \sign(\dest)\frac{\Big(\vert \dest \vert -   A \sqrt{\frac{\sest}{N}}\Big)_+}{\lambda \sest} = \left(\bfan - \sign(\dest)\frac{A \sqrt{\frac{\sest}{N}}}{\lambda \sest}\right)\myind{\dest^2 \geq A^2\frac{\sest}{N}},
    \end{align*}
    where $A := \phi(\Phi^{-1}(\alpha))/\alpha$ with $\phi(\cdot)$ denoting the probability density function and $\Phi^{-1}(\cdot)$ the quantile function of the standard normal distribution. This results, under the assumption that $\sest \equiv \sigma^2$, in an explicit solution $f$ of the out-of-sample performance
    $
        \entr_\lambda\left( \bfarr \Delta S_{1}\right) 
        = f(\mu, \sigma, \lambda, \alpha, N).
    $
\end{restatable}
\begin{proof}
    See \Cref{app:Iid}.
\end{proof}

\begin{figure}
    \centering
    \hspace{1cm}
    \includegraphics[height=4cm]{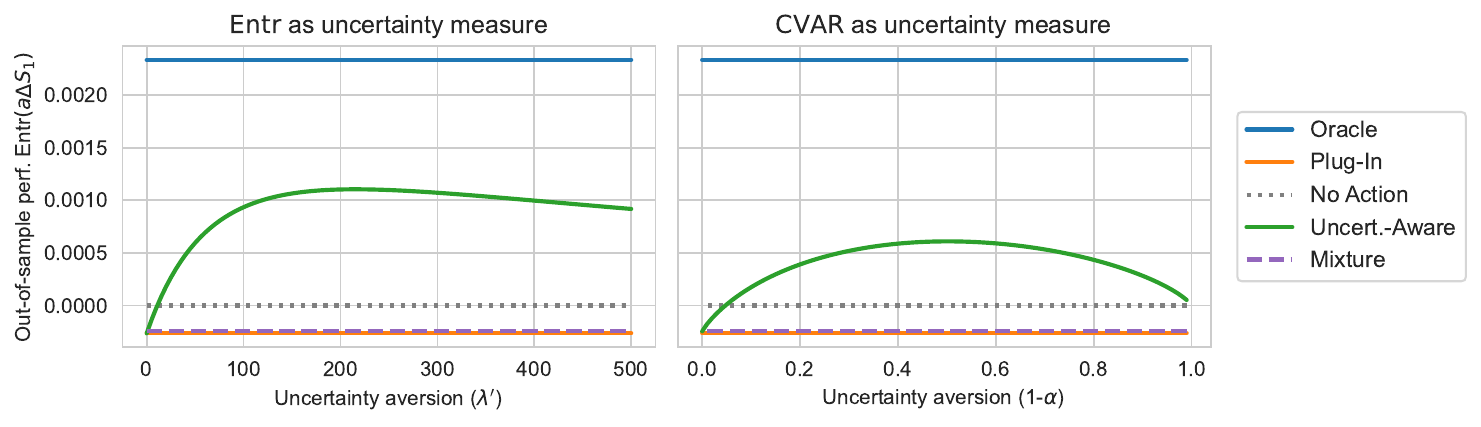}
    \caption{The out of sample performance of the optimal, plug-in, and mixture distribution (as defined later in \Cref{sec:ComparingAnalyticallyIIDNormal}), as well as uncertainty-aware strategies, using $\entr$ and $\cvar$ as uncertainty measures, \change{respectively. Note that mixture and plug-in deliver negative out-of-sample performance in this example.}}
    \label{fig:AOOSP}
\end{figure}

The difference in strategies is best exemplified when considered for a choice of parameters. Here, let $\lambda = 0.84$, $N = 140$, $\sigma = 0.2/\sqrt{255}$, and $\mu = 0.2/255$. 
The performance of the strategies is shown for varying choices in uncertainty aversions $\lambda'$ and $\alpha$, respectively, in \Cref{fig:AOOSP}.

The out-of-sample performance of different strategies, illustrated in \Cref{fig:AOOSP}, provides a clear comparison between plug-in and uncertainty-aware decision-making.  The optimal strategy, which assumes perfect knowledge of parameters, unsurprisingly achieves the highest performance. 

A key benchmark when evaluating performance is whether a strategy achieves an out-of-sample performance greater than zero, which would indicate that it outperforms the trivial (zero-investment) strategy. Notably, the plug-in strategy suffers from the estimation error so drastically that it is less effective than the trivial strategy.

Among the uncertainty-aware approaches, both the entropic risk-based strategy and the $\cvar$-based strategy exhibit robustness against model uncertainty, for sufficient robustification preferences. 
We see that both strategies converge for small uncertainty aversion to the plug-in strategy, and oppositely to the trivial strategy for very high uncertainty aversion. \oldchange{This is stated in the corollary below.

\begin{restatable}{cor}{UAStratLimits}\label{cor:UAStratLimits}
For the uncertainty-aware policy \eqref{id:UAEntrEntr} holds
    \begin{align*}
        \lim_{\lambda'\to 0}\bfar = \bfan, \quad 
        \lim_{\lambda'\to \infty}\bfar \equiv 0, \quad 
        \lim_{N \to \infty}\bfar = \bfao.
    \end{align*}
Analogously, for the uncertainty-aware policy \eqref{id:UACvarEntr} holds
    \begin{align*}
        \lim_{\alpha \to 1}\bfarr = \bfan, \quad 
        \lim_{\alpha \to 0}\bfarr \equiv 0, \quad 
        \lim_{N \to \infty}\bfarr = \bfao.
    \end{align*}
\end{restatable}
\begin{proof}
    See \Cref{app:Iid}.
\end{proof}
}
\begin{figure}
    \hspace{2.25cm}
    \includegraphics[height=3.8cm]{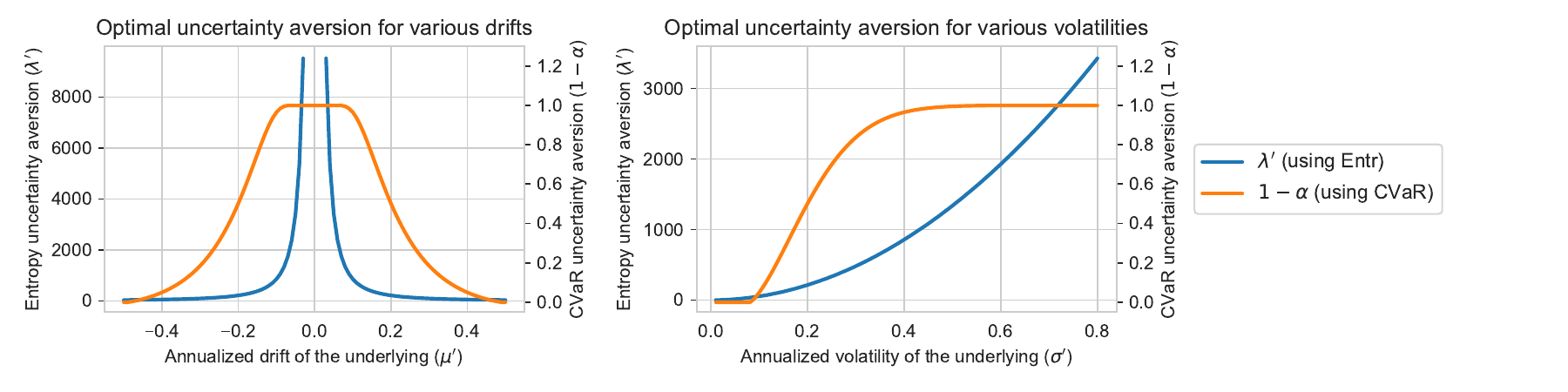}
    \caption{The optimal choice of uncertainty aversion $\lambda', \alpha$ that result in the highest out-of-sample performance for the uncertainty-aware strategies \eqref{id:UAEntrEntr} and \eqref{id:UACvarEntr}, for various testing environments defined by annualized drift and volatility $\mu', \sigma'$. If the parameter is not varied, it corresponds to $\mu'=0.2$ and $\sigma'=0.2$ as in the example above. It is evident that markets with lower drifts compared to volatilities require higher robustification.}
    \label{fig:AOOSPOptUA}
\end{figure}




\oldchange{
When examining the uncertainty–aversion parameters that yield optimal robustification, several systematic patterns emerge (\Cref{fig:AOOSPOptUA}). First, larger drift magnitudes \(\mu\)—which correspond to stronger predictive signals—require relatively mild uncertainty–aversion (smaller \(\lambda'\) or larger \(\alpha\)), consistent with intuition that clear trends demand less protection against model misspecification. Conversely, when the drift is weak, the optimal robustification converges to its extreme: \(\lambda'\to\infty\) (for the entropic penalty) and \(\alpha\to 0\) (for the $\cvar$ penalty), indicating that the agent effectively defaults to the trivial (zero–investment) strategy in the absence of a reliable signal. Finally, varying the volatility parameter produces the inverse relationship: high volatility (noisy returns) calls for stronger robustification, whereas low volatility (sharp signals) allows for more aggressive positioning with reduced uncertainty–aversion. 
}

\paragraph{Comparing to the mixture of measures approach.}


A natural question arises when considering uncertainty-aware strategies: if we already have a distribution over possible models, why not simply operate under the mixture measure (see \cite[Definition 6.1]{CuchieroGazzaniKlein22} and references therein).

While this approach incorporates model uncertainty to some extent, it does so in a way that fundamentally differs from uncertainty-aware strategies. Specifically, it averages over possible models without explicitly penalizing or adjusting for the risk associated with model misspecification. In the context of investment decisions, this means that the joint measure strategy seeks to optimize performance under the averaged distribution of returns, rather than explicitly accounting for worst-case scenarios or tail risks. 
To assess whether robustification via explicit uncertainty measures offers advantages over the mixture approach, we analytically derive the corresponding strategy and its out-of-sample performance.

The \textit{mixture} strategy is defined by\footnote{Note that if $J$ is linear in the measure component, $\bfac$ coincides with $\bfar$ when choosing the expectation operator as uncertainty measure. Linearity of $J$ is however not generally the case and atypical in applications.}
    \begin{align}
        \bfac 
        := 
        \argmax_{\Strategy \in \StrategySpace} 
        \Objective\left(X(\Strategy), \int\bbP_\theta \ModelDist(d\theta) \right) \label{id:Mixture},
\end{align}
which, with the above choices (and \Cref{lem:SolveEntr}) reduces to
\begin{align}
        \bfac = 
        \argmax_{\Strategy \in \R} 
        \left\{
        \entr_\lambda(a \Delta S_1)
        \middle|
        \Delta S_1 \sim \calN(\dest, \sigma^2 + \sigma^2/N)
        \right\} 
         = 
        \frac{\dest N}{\lambda\sigma^2 (N + 1)}.
        \label{id:MixtureIid}
\end{align}
With this, we can compute its out-of-sample performance (see \Cref{lem:OOSPJoint})
\begin{align*}
    \entr_\lambda(\bfac\Delta S_{1}) =
    \frac{\mu^2N}{\lambda\sigma^2 (N + 1)} 
    -
    \frac{N(\mu^2 + \sigma^2)}{2\lambda\sigma^2 (N + 1)^2}.
\end{align*}
As we see in \Cref{fig:AOOSP}, while the mixture approach offers some degree of robustness, it is ultimately insufficient in this setting. This highlights the necessity of structured uncertainty quantification beyond mere model averaging.

\oldchange{
The reason the mixture‐measure strategy largely coincides with the plug-in strategy is that the latter is a trivial function of the mean. Therefore, the mixture approach averages over an estimator of the plug-in strategy, resulting in very close results. We expect starker contrasts in scenarios where the optimal strategy is a non-trivial function of the mean.
}

\begin{figure}
    \centering
    \hspace{2cm}
    \includegraphics[height=4.5cm]{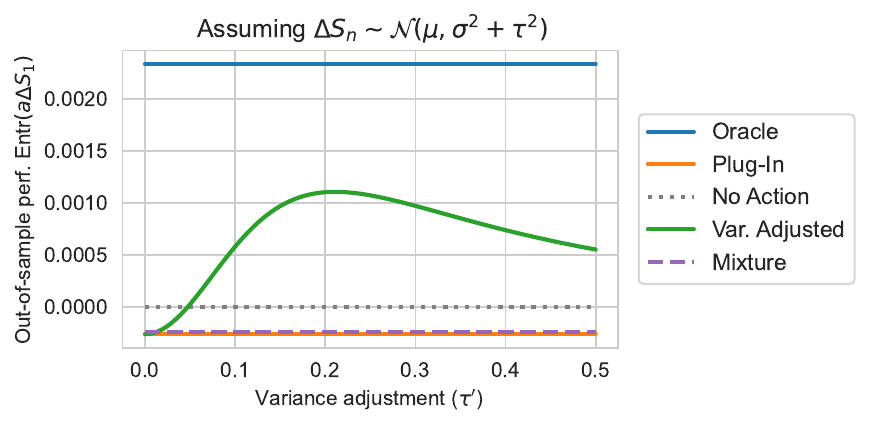}
    \caption{Performance of the plug-in strategy with a custom variance adjustment $\tau^2$ in context of \Cref{exa:Plug-inNormal}.}
    \label{fig:ManualVolAdjustment}
\end{figure}

Note that this strategy simply amounts to the plug-in strategy if we estimated $\calN(\dest, \sigma^2 + \sigma^2/N)$ as a distribution for the next return. 
\oldchange{
This means, the mixture strategy in this example can be written as the following \textit{simple variance adjustment} strategy
\begin{align*}
        \bfava = 
        \argmax_{\Strategy \in \R} 
        \left\{
        \entr_\lambda(a \Delta S_1)
        \middle|
        \Delta S_1 \sim \calN(\dest, \sigma^2 + \tau^2)
        \right\} 
\end{align*}
for choosing the variance adjustment $\tau^2 = \sigma^2/N$. Clearly, the adjustment $\tau^2 = \sigma^2/N$ is too little, offering the question: which adjustment \textit{would} result in a meaningful performance improvement? In \Cref{fig:ManualVolAdjustment} we plotted the performance of $\bfava$ for varying adjustments $\tau^2$, we see that the adjustments for sufficient robustification are relatively high and there is no trivial relation to the de facto uncertainty of the estimation of $\dest$.
}

\oldchange{Upscaling of the variance to account for uncertainty has been studied extensively in the literature, especially in the context of improving estimators for risk estimation, see for example \cite{pitera2023novel} and references therein.}

\begin{rem}[Relation to the uncertainty-aware strategy]

The mixture approach coincides with the uncertainty-aware strategy when the objective $J$ is linear in its second argument, and if the uncertainty measure is the expectation: indeed, if $\Objective(X(\Strategy), \int\bbP_\theta \ModelDist(d\theta)) = \int \Objective(X(\Strategy),\bbP_\theta)  \ModelDist(d\theta)$,
then
\begin{align*}
    \argmax_{\Strategy \in \StrategySpace} 
    \left\{
    \UM\big(\Objective(X(\Strategy), \bbP_\theta), \ModelDist\big) 
    \right\}
&=  \argmax_{\Strategy \in \StrategySpace} 
    \left\{
    \int_\Theta \Objective(X(\Strategy),  \bbP_\theta)
    \,  \ModelDist(d\theta) 
    \right\} \\
    &=  \argmax_{\Strategy \in \StrategySpace} 
    \left\{\Objective(X(\Strategy), \int\bbP_\theta \ModelDist(d\theta)) 
    \right\} 
\end{align*}
which  coincides with the mixture strategy. 

\end{rem}

\section{Subsampling: An Agnostic Choice of Model Distribution}\label{sec:SubSampling}

In many practical applications, obtaining clear and reliable distributions for model parameters is challenging. In these cases, it is beneficial to adopt methods that yield reasonable approximations of the model distribution, and we propose the use of subsampling for this purpose. \oldchange{The method is a simple and computationally efficient way to derive a distribution $\frakP$ from a single estimated model $\hat\bbP$. Although limited in generality, this method is practically attractive: it is easy to implement, requires minimal hyperparameter tuning (which is often delicate in other approaches), and integrates seamlessly with existing modelling pipelines.}

Importantly, we argue that when uncertainty measures are applied, the precision of the model distribution becomes less critical than for the mixture method; even with this ad hoc approach, substantial performance gains can still be achieved.

We proceed by defining subsampling in this context. Generally, for estimated measures, closed-form solutions to the investment strategies \eqref{id:Plug-inStrategy} are unavailable, and one must resort to Monte Carlo methods.  \oldchange{Recall that previously the payoff \(X(a)\) was considered in a general form, involving both the action and the evolution of the asset. However, in most cases the evolution of the asset is not influenced by the action and can be treated separately. In this section we will adopt this viewpoint, i.e.\ assume that the payoff is given as a function of the action $a$ and some underlying $S$,
\begin{align} 
    X(a) = f(a,S).\label{id:FunctionalPayoff}
\end{align}
It suffices therefore in the following to consider (the unknown) $\bbP$ as the distribution of $S$ itself.}

An \emph{$n$-subsample}, with $n \in \N$, is  simply a set of events $(\omega_i)_{i=1}^n \subseteq \Omega$ drawn from the estimated distribution $\hat \bbP$, which results in the empirical evaluation 
\[
\bigl( S( \omega_i) \bigr)_{i=1}^n.
\]
As used in Monte-Carlo style methods, we obtain the \textit{subsample measure} by weighting each $\omega_i$ equally, i.e. 
%
\begin{align}
\bbP_\theta := \mathcal{U}\bigl((\omega_i)_{i=1}^n\bigr) = \frac 1 n \sum_{i=1}^n \delta_{\omega_i},\label{id:subsampling}
\end{align}
where $\delta_\omega$ denotes the Dirac measure at $\omega$ and $\mathcal{U}$ refers to the uniform distribution of the events $\theta := (\omega_i)_{i=1}^n$. \oldchange{Formally this can be cast into the setting of uncertainty-aware strategies of this paper by setting $\ParameterSpace := \Omega^n$ and $\ModelDist = \hat \bbP^n$.}\footnote{\oldchange{Note that this also applies to general payoffs $X$
not necessarily of the form \eqref{id:FunctionalPayoff}. However, since this case is considerably more abstract and of limited practical relevance, we do not pursue it further here.}}



\paragraph{Connection to the uncertainty in the normal i.i.d.~example.}
In the setting of \Cref{sec:ComparingAnalyticallyIIDNormal}, the subsampling technique introduced above leads \change{in the limit} to the same strategies as in  \Cref{lem:OospUAIidEntrEntr} and \Cref{lem:OospUAIidEntrCVaR}, respectively.
Indeed,  first recall that \oldchange{we assumed} the agent estimates \oldchange{for the plug-in strategy} that $\hat \bbP$ such that $\Delta \pS_1 \sim \calN(\dest, \sest) \approx \calN(\dest, \sigma^2)$. 
 \oldchange{We will now apply the above subsampling technique to obtain $\ModelDist$.} The $N$ samples from $\hat \bbP$ have distribution $(\Delta \pS_{1, i})_{i = 1}^N \sim \calN(\dest, \sigma^2)^N$. \change{Of course the actual mean estimate per subsample will be different and distributed normally with mean and variance}
\begin{align*}
    \myE{\hat \bbP}{\frac{1}{N} \sum_{i = 1}^N\Delta \pS_{1, i}} = \dest,  
    \quad\text{and}\quad
    \myV{\hat{\bbP}}{\frac{1}{N}\sum_{i=1}^N \Delta \pS_{1, i}} =\frac{\sigma^2}{N}.
\end{align*}
(We again argue that the variance estimation is sufficiently accurate.)
In the context of \Cref{exa:Plug-inNormal,exa:RobustNormal} this means, that if we use the central limit theorem (and \eqref{id:MVisENT}) to approximate $\entr_\lambda(\Delta S_1) \approx \mv_\lambda(\Delta S_1)$ for $\Delta S_1 \sim \calU((\Delta \pS_{1, i})_{i = 1}^N)$, the strategy resulting from our subsampling technique have\ \change{the same 
theoretical out-of-sample performances as their analytical versions} \Cref{lem:OospUAIidEntrEntr,lem:OospUAIidEntrCVaR}.

\begin{figure}[p]
    \centering
    \includegraphics[height=4.5cm]{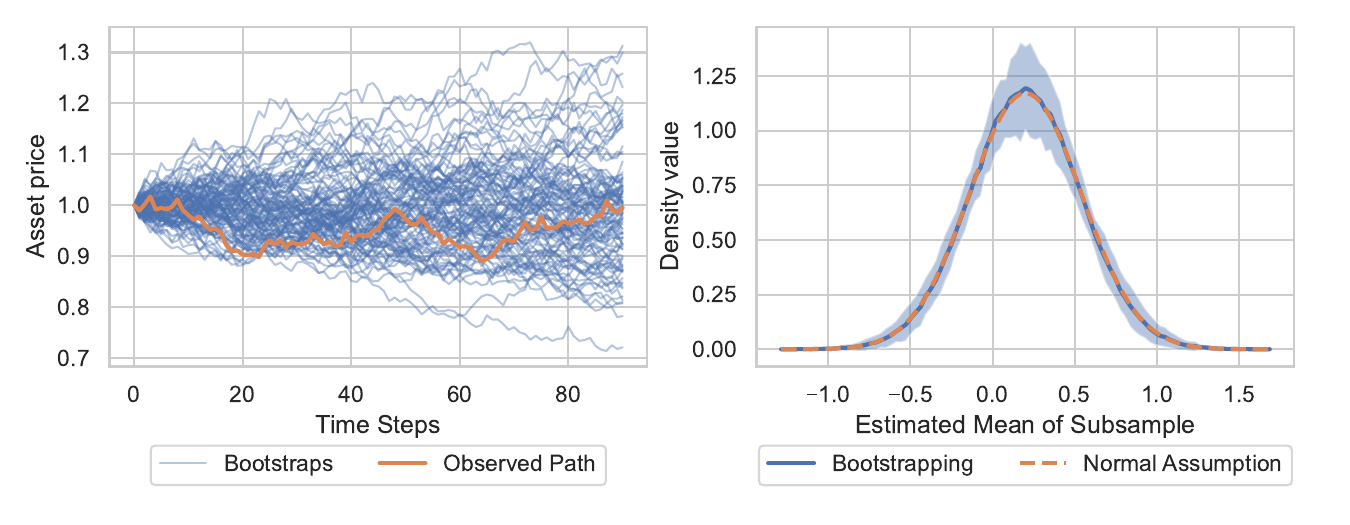}
    \caption{On the left, one path is ``bootstrapped'' by sampling its increments (with repetition) to create new paths. On the right, mean estimations based on bootstrapping are depicted and compared to $\smash{\mathcal{N}(\dest, \sest/N)}$, the mean estimations obtained from subsampling with a normal distribution. The solid blue line represents the average approximation, and the shaded area represents one standard deviation.}
    \label{fig:Bootstrap}
\end{figure}
\begin{figure}
    \centering
    \hspace{3cm}\includegraphics[height=4.5cm]{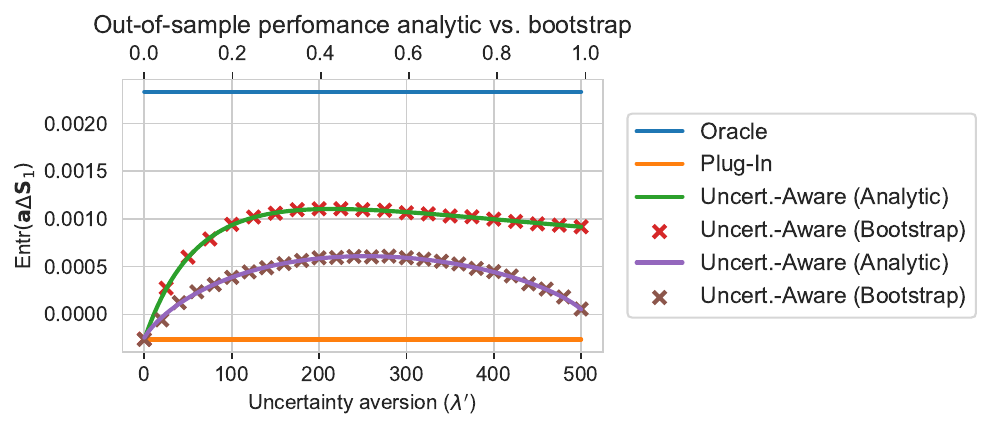}
    \caption{Comparing out-of-sample performance of the analytic strategies (\Cref{lem:OospUAIidEntrEntr,lem:OospUAIidEntrCVaR}) to bootstrapping \change{increments} (which is equivalent to \change{i.i.d.~}subsampling under the empirical measure). It is apparent that the approximation showed \Cref{fig:Bootstrap} is good enough (on average) to obtain negligible differences between both approaches.}
    \label{fig:BootstrapToAnalytic}
\end{figure}
\paragraph{Bootstrapping vs subsampling.} 
In the example above, the agent estimated a normal distribution to then drew samples from this distribution. Bootstrapping, as it is used by practitioners in investing contexts, on the other hand operates directly under the empirical measure of the observed returns, by randomly drawing (with replacement) from the set of observed returns. \change{For the one-step case this formally corresponds to subsampling with} $\hat \bbP = \calU((\Delta \pS_{t})_{t = -N+1}^0)$. 
\change{This is simpler than first estimating a more complex model as long as i.i.d.~returns of a stock are considered, but as we see below sub-sampling lends itself better to data with time dependencies.}

\oldchange{
\change{A~common use case for bootstrapping is to compose paths of several steps in the future by adding up historic returns drawn with repitition randomly,} as vizualized in the left-hand-side of \Cref{fig:Bootstrap}. This means, if $(\omega)_{i=1}^N \sim \hat \bbP^n$ then there exists for every $i\in \myset{1}{N}$ and $j\in \myset{1}{N}$, such that $\Delta S_1(\omega_i) = \Delta S_{-j}$. Note that this corresponds exactly to the definition of an $N$-subsample drawn from $\hat \bbP$.
}

Comparing the mean-estimation of subsampling after a normal assumption, with the one resulting from bootstrapping, shows how close those approaches are (see right-hand-side of \Cref{fig:Bootstrap}). Indeed, the central limit theorem guarantees that they coincide for large enough samples $N$. \oldchange{As one can see in \Cref{fig:BootstrapToAnalytic}, the changes in performance are marginal and practically negligible.}

\begin{rem}[Considerations for the estimated measure $\hat \bbP$]
When $\hat \bbP$ is taken as the empirical measure—as in naive bootstrapping—resamples are limited to the $N$ observed data points, which can severely underrepresent tail risk and lead to unreliable CVaR estimates (\Cref{fig:Bootstrap}). Smoothing the center \cite{wand1995kernel} and thickening the tails \cite{coles2001introduction,mcneil2000estimation,pitera2022estimating} can mitigate this, though such refinements are beyond the scope of this work.
\end{rem}

\change{
The effectiveness of any subsampling (or bootstrap) scheme hinges on how well the underlying estimated data-generating measure $\hat \bbP$ captures the salient features of the market.  Since bootstrapping inherently assumes that returns are i.i.d., it removes any path dependencies such as auto-correlation in volatilities. Our sub-sampling approach improves upon this simpler method by allowing to first draw a model which captures relevant features of the model and then use our uncertainty approach to robustify it.}


\begin{figure}[p]
    \centering
    \includegraphics[height=9cm]{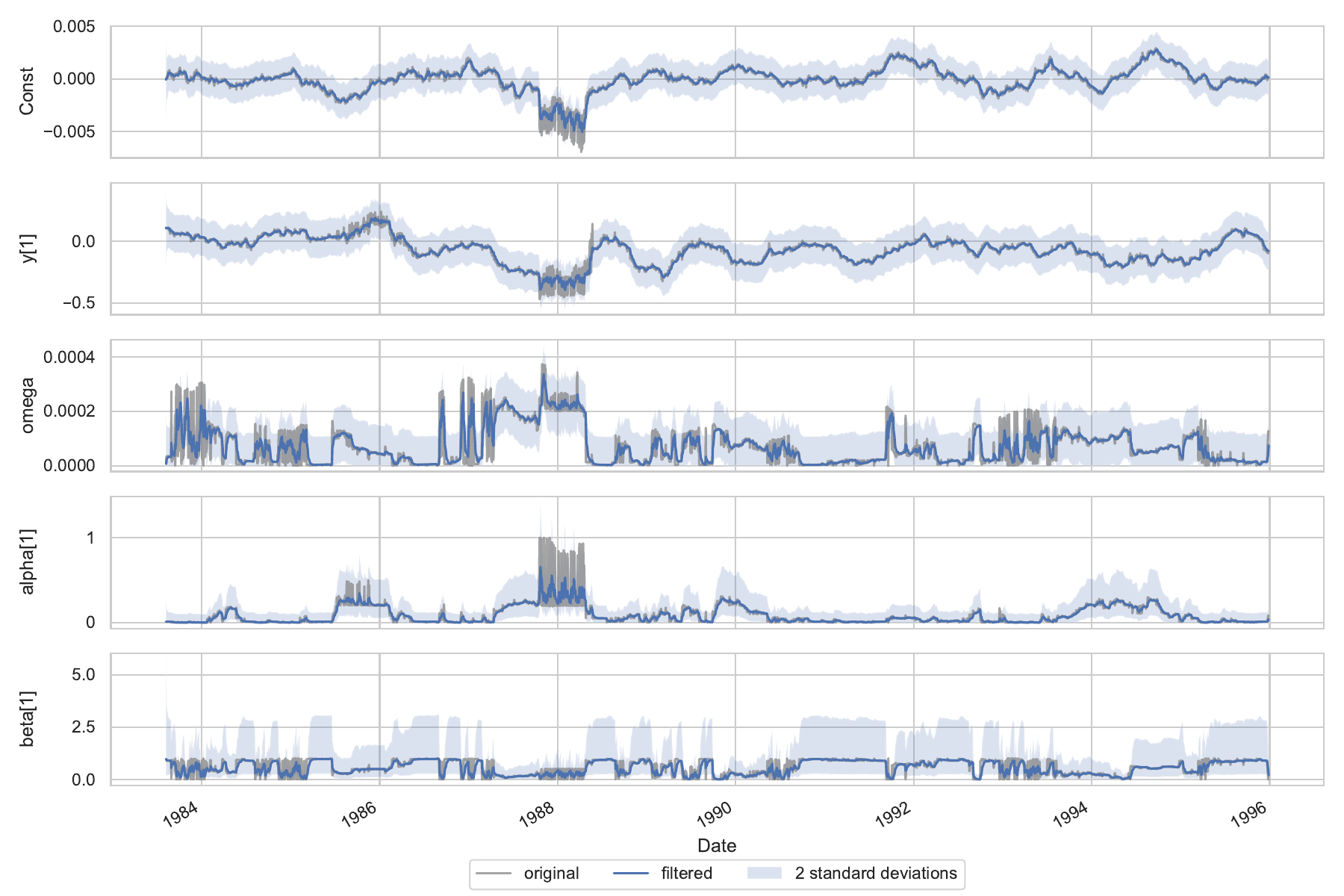}
    \caption{ARCH(1)-GARCH(1,1) parameters calibrated to PEP-KO pair process with Kalman filter uncertainty eastimation.}
    \label{fig:KalmanFiltering}
\end{figure}
\begin{figure}[p]
    \centering
    \includegraphics[height=5cm]{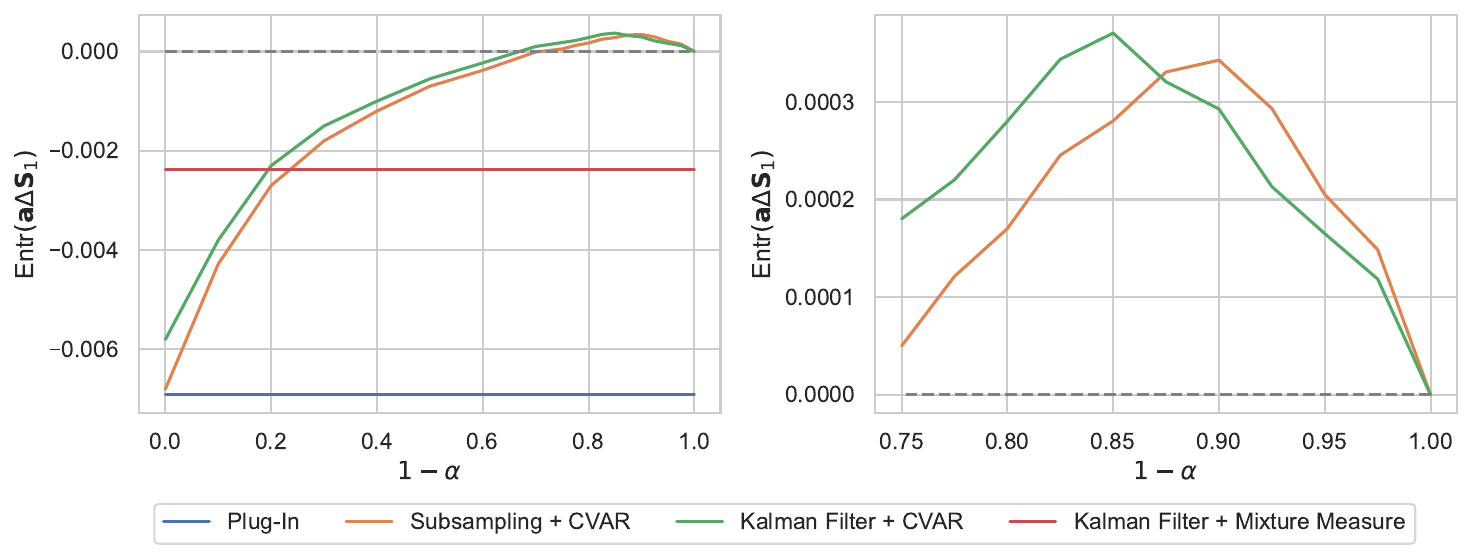}
    \caption{Out-of-sample performance of strategies on PEP-KO pair process. Here, ``Subsampling + CVaR'' uses a subsample size of $30$ and $\cvar_\alpha$ as uncertainty measure; ``Kalman Filter + CVaR'' uses the same uncertainty measure, uses a Kalman Filter to obtain a distribution of AR(1)-GARCH(1,1) models; and ``Kalman Filter + Mixture Method'' employs the mixture strategy \eqref{id:Mixture} on this distribution. It is clear that Kalman filtering has a positive impact on the performance of the strategy, most notably when paired with $\cvar$ as uncertainty measure. Note, however, that similar performance gains can be achieved with the subsampling method. }
    \label{fig:SubsVSKF}
\end{figure}

\paragraph{Example on Real Market Data using ARCH/GARCH models.}
To evaluate the practical performance of our subsampling approach, we examine a pairs-trading task between PepsiCo Inc. (PEP) and Coca-Cola Co. (KO) using daily data from 1983–1996\footnote{This period is deliberately chosen, as the pair signal became weaker with the diversification of PEP in later years \cite{GatevGoetzmannRouwenhorst06}.}. \change{The idea of this section is to back-test the plug-in and uncertainty-aware strategies over this period. This means, we compute the strategies for each day, based on the recent history of the market and invest accordingly. Eventually, we will compare the performance of the strategies over the 13-year period. }

\oldchange{Firstly, let us define the plug-in strategy that we will subsequently robustify.} In this setting, to exploit the mean-reversion property of the pairs process for maximizing \(\entr_{0.87}\), we require \oldchange{a model $\hat \bbP$} that not only captures the mean-reversion but also provides reliable volatility estimates. Accordingly, we model the returns of the spread process 
\(S\) using an autoregressive model for the mean and a GARCH(1,1) process for the conditional variance. 
\oldchange{Under $\hat \bbP$, we therefore assume that} the {return} equation is given by
\[
\Delta S_t = \hat c + \hat y \Delta S_{t-1} + \epsilon_t,\quad \epsilon_t = \sigma_t Z_t,
\]
where \(\hat c\in\mathbb{R}\) is the intercept, \(\hat y\in\mathbb{R}\) the autoregressive coefficient, and \(Z\) is an i.i.d. sequence with zero mean and unit variance. The conditional variance \oldchange{under $\hat \bbP$} follows the GARCH(1,1) dynamics:
\[
\sigma_t^2 = \hat \omega + \hat\alpha \epsilon_{t-1}^2 + \hat\beta \sigma_{t-1}^2,
\]
with \(\hat \omega>0\) representing the long-run variance, \(\hat \alpha\geq 0\) the impact of past shocks, and \(\hat\beta\geq 0\) the persistence of volatility.

Assuming that a history of the most recent 150 trading days \oldchange{is sufficiently informative} for estimating the mean-reversion properties, we fit the simple AR(1)-GARCH(1,1) model \oldchange{for every day over the 13-year time period. Therefore, we have an estimated measure $\hat\bbP$ for each of those days, which we can use to compute the plug-in strategy, according to \eqref{id:Plug-inStrategy}.}

We can now follow \oldchange{the three steps detailed in  \Cref{sec:uncertainty aware strategy}} to obtain an uncertainty-aware strategy. 
\oldchange{
We \begin{enumerate*}[label=(\roman*)]
    \item choose the class of models $(\bbP_\theta)_{\theta \in \Theta}$ as all possible subsamples as in \eqref{id:subsampling},
    \item use the subsampling method to obtain the distribution $\frakP$, and 
    \item choose an uncertainty measure, here $\cvar$ with uncertainty aversion $\alpha \in [0,1]$.
\end{enumerate*}
}
Effectively, we use the estimated measure $\hat \bbP$ (this is the ARCH-GARCH model), to (repeatedly) create $n$-subsamples by simulating $\Delta S_1$ $n$-times as described by the ARCH-GACH dynamics. These $n$ simulations represent a measure ${\bbP_\theta}\sim\ModelDist$. With this, we can follow \eqref{id:RobustStrategy} to compute the uncertainty-aware strategy.

As illustrated in \Cref{fig:SubsVSKF}, the uncertainty-aware strategy, with a subsample size of \oldchange{$n= 30$}, significantly improves performance by achieving a maximum at \(\alpha \approx 15\%\) and yielding positive utility---surpassing even the performance of the trivial approach.

For comparison, we employ a Bayesian approach using a simple Kalman Filter (see e.g. \cite{chui2017kalman} for details on Kalman filtering and implementation) fitted to daily model parameter estimations. In this implementation, the parameters \(\omega\), \(\alpha_1\), and \(\beta_1\) are transformed via a log-transform prior to ensure non-negativity. The resulting filtered process, along with its confidence region, is displayed in \Cref{fig:KalmanFiltering}.

When applying the uncertainty measure approach (using \(\cvar\)) to the distribution of AR(1)-GARCH(1,1) models obtained by filtering, we observe a performance improvement that is remarkably similar to that achieved by the subsampling approach, with the Bayesian method slightly outperforming for most choices of the uncertainty aversion parameter \(\alpha\), see \Cref{fig:SubsVSKF}. Importantly, however, these gains are marginal, 
and might be outweighed by the simplicity and practicality of the subsampling approach.  

Furthermore, while the robustification obtained via the mixture method once again enhances the performance relative to the uncertainty-unaware strategy, it fails to provide sufficient improvement to outperform the trivial (zero-investment) strategy and fails to compete with a (reasonably calibrated) strategy that utilizes uncertainty measures.


\section{Overcoming memory constraints: A CVaR-stochastic gradient allowing deep learning applications}\label{sec:CVaRGradientDescent}

\oldchange{
A major computational challenge in applying the uncertainty-aware strategies of this article is the memory burden associated with simulating paths under a distribution of models. Sampling multiple models from a distribution $\frakP$ and simulating returns/paths for each of these models leads to a quadratic computational cost. While modern GPUs can mitigate the processing time by effective parallelization, their memory remains a bottleneck, limiting the scale and complexity of feasible models.

This becomes especially problematic when considering high-dimensional problems or when transitioning from simple single-period return predictions to more involved tasks that require simulating entire asset price paths. These tasks inherently involve larger state spaces and require repeated evaluations across many model variants—quickly overwhelming memory resources.

To address this issue, we propose a memory-efficient variant of stochastic gradient descent (SGD) tailored to uncertainty-aware optimization with Conditional Value-at-Risk (CVaR). Crucially, this method operates by simulating paths under a single model at a time rather than in bulk. This design significantly reduces memory consumption while still enabling model risk to be captured effectively. Moreover, the independence of individual simulations allows for seamless distribution across multiple compute nodes, making it well-suited for modern deep learning pipelines.
}

\oldchange{
Before introducing our adapted optimization method, we briefly recall the following result: SGD, when applied to the plug-in objective $\Objective(X(a), \bbP_\theta)$ with $\bbP_\theta$ the uniform distribution over a mini-batch $\theta := (\omega_i)_{i=1}^n \sim \hat\bbP^n$, converges---under standard regularity conditions (see \cite[Theorem 14.8]{ShalevBen14})---towards the expectation of the objective under the sampling distribution,
\[
\mathbb{E}_{\theta \sim \hat\bbP^n}\left[\Objective(X(a), \bbP_\theta)\right].
\]
%
%
%
%

Clearly, if the objective $J$ is unbiased under empirical evaluation, the mini-batch gradient descent simply optimizes the objective under the measure  $\hat\bbP$. This implies that the standard mini-batch approach inherently accounts for model uncertainty, albeit implicitly, by using the expectation operator as the underlying measure of uncertainty.}

To explicitly incorporate model risk into the optimization process, we propose an adaptation of the gradient descent procedure, replacing the expectation operator with the $\cvar$ as the uncertainty measure. This results in an alternative stochastic gradient descent method (\Cref{alg:cvar_sgd}) that prioritizes performance under adverse scenarios, thereby enhancing robustness in uncertain environments.

\begin{algorithm}[t]
\small
\caption{$\cvar$ Stochastic Gradient Descent}
\label{alg:cvar_sgd}
\SetAlgoLined
\DontPrintSemicolon
\KwIn{Initialize $a^{(0)}$ with parameters $\phi^{(0)}$, learning rate $\eta$, batch size $m$, $\cvar$ level $\alpha$}
\KwOut{Optimized parameter $\phi^*$}
$t \gets 0$ \;
\While{not converged}{
    \oldchange{Initialize ``worst objective'' and ``worst gradient'' sets $\mathcal{J} \gets \emptyset$, $\mathcal{G} \gets \emptyset$} \;
    Independently sample integer $k$ such that $\mathbb{E}[k/m] = 1-\alpha$ \;
    \For{$i \gets 1$ \KwTo $m$}{
        Sample a measure $\mathbb{P}_i \sim \ModelDist$ \;
        Compute objective $J \gets  \Objective\big(X(a^{(t)}); \mathbb{P}_i\big)$ 
        \oldchange{
        \If{$\mathcal{J}$ has less than $k$ entries}{
            Insert $J$ to $\mathcal{J}$, \change{compute gradient $g \gets \nabla_\phi J_i$ and add to} $\mathcal{G}$ \;
        }
        \If{$\mathcal{J}$ has $k$ entries \textbf{and} $J$ is smaller than the maximum of $\mathcal{J}$}{
            Replace $J$ with the maximum of $\mathcal{J}$, \change{compute gradient $g \gets \nabla_\phi J_i$} and replace the corresponding element in $\mathcal{G}$ with $g$\;
        }}
    }
    Update $a^{(t)}$ to $a^{(t+1)}$ with parameters $\phi^{(t+1)}$:
    \[
        \phi^{(t+1)} \gets \phi^{(t)} - \eta\, \frac{1}{k} \sum_{g \in \mathcal{G}} g
    \]
    $t \gets t + 1$ \;
}
\end{algorithm}




Notably, the proposed algorithm does not require the simultaneous computation of multiple objectives and gradients, thereby maintaining memory requirements comparable to a $(1-\alpha)$-fraction of the naive stochastic gradient descent, as at most $(1-\alpha)\cdot m$ gradients need to be saved simultaneously in memory. This reduction can be crucial, especially for a high-parametric class of strategies $\StrategySpace$.

Despite this reduced memory footprint, the algorithm remains computationally efficient. Since the gradient computations for different sampled measures \(\mathbb{P}_i\) are entirely independent, they can be executed in parallel across multiple processors or even distributed across different machines. This parallelism enables significant speedups, especially for large-scale models.

From a theoretical perspective, the key property of the \(\cvar\)-SGD algorithm is that, in expectation, it discards the gradients corresponding to the best-performing \(\alpha\)-fraction of objectives. Consequently, the computed update direction serves as an unbiased estimator of the gradient of the uncertainty-aware \(\cvar\) objective. This leads to the following result:

\begin{restatable}{thm}{CVARSDGTHM}[$\cvar$-SGD]\label{thm:CVARSGD-randomk}
Let $\Phi = \{ \phi \in \mathbb{R}^d : \|\phi\| \leq B \}$ for some $B > 0$, and assume that for every $\bbP \sim \ModelDist$, the function $\phi \mapsto J(X(a_\phi), \bbP)$ is convex and $\rho$-Lipschitz, i.e.,
$
\left\| \nabla_\phi J(X(a_\phi), \bbP) \right\| \leq \rho 
$ $\ModelDist$-almost surely.
Let $\alpha \in (0,1)$ and define the objective
$
\mathcal{L}(\phi) := \cvar_{\alpha}\left( J(X(a_\phi), \bbP) \,\middle|\, \bbP \sim \ModelDist \right).
$
Let $(\phi_t)_{t \geq 0}$ be the sequence generated by \Cref{alg:cvar_sgd}, using step size $\eta_t = {B}({\rho \sqrt{t}})^{-1}$, and define the average iterate
$
\bar{\phi}_t := \sum_{s=1}^{t} \phi_s /t.
$
Then, for all $t \geq 1$,
\[
\mathbb{E}[\mathcal{L}(\bar{\phi}_t)] - \min_{\phi \in \Phi} \mathcal{L}(\phi) \leq \frac{B \rho}{\sqrt{t}}.
\]
\end{restatable}

\subsection{Improved precision in high-dimensional investing}\label{sec:HighDim}

Firstly, we want to show the reduction of the memory constraints by $\cvar$-SGD in a controlled environment. For this, we consider a high-dimensional version of the standard investing example, \Cref{exa:Plug-inNormal}. By increasing the dimensionality, not only the simulations become more expensive, but it also makes the strategy space $\StrategySpace$ richer and therefore the problem harder. The idea is to first derive an analytic baseline, which we then approximate in a realisitc setting with and without $\cvar$-SGD for varying $\alpha \in [0,1]$ and analyze the algorithms effect on precision.


Formally speaking, we work in a multi‐dimensional version of the Gaussian setup introduced in \Cref{sec:ComparingAnalyticallyIIDNormal}.  Let $d \in \N$,
and $(\Delta S_t)$ be i.i.d.\ with
\[
\Delta S_t \;\sim\;\mathcal N_d(\mu,\Sigma),
\qquad t=-N+1,\dots,0,
\]
where $\mu\in \R^d$ is the true mean vector and $\Sigma \in \R^{d\times d}$ is the true covariance matrix
(both unknown). 
Then,  the sample mean
\[
\dest \;=\;\frac{1}{N}\sum_{t=-N+1}^0\Delta S_t,
\]
has distribution \(\hat\mu\sim\mathcal N_d(\mu,\Sigma/N)\). With similar arguments as in \Cref{sec:ComparingAnalyticallyIIDNormal}, we neglect the estimation error of \(\Sigma\) and focus on the estimation error in \(\dest\).

The agent's actions are now multi-dimensional, i.e. $a \in \R^d =: \StrategySpace$. Analoguously to \Cref{exa:Plug-inNormal}, the payoff of 
{the considered} contingent claim will be the return scaled by the corresponding action, i.e. $X(a) := a^\intercal \Delta S_1$. Again, we choose the agent has an entropic risk objective with risk aversion $\lambda > 0$. For completeness, we can show (\Cref{lem:HighDimObjective}) that the plug-in objective results into the following mean-variance objective
\[
J(X(a), \hat\bbP)\;=\;a^\intercal\dest \;-\;\frac{\lambda}{2}\,a^\intercal\Sigma\,a.
\]
which is optimized by (see \Cref{lem:plug-in_hd}) $\bfan =\,\Sigma^{-1}\, \dest / {\lambda}$.

\paragraph{Analytic solutions.} We now want to robustify this strategy using $\cvar_\alpha$ for some $\alpha \in [0,1]$. As mentioned above, we will start with some analytic considerations, along the line of \Cref{sec:ComparingAnalyticallyIIDNormal}, to compare the later approximations against.

\begin{restatable}{lemma}{MDCVARStrategy}
\label{lem:MDCVARStrategy}
The $\cvar$ uncertainty-aware strategy, which is analogue to \eqref{id:UACvarEntr}, and defined by
    \begin{align*}
    \argmax_{\Strategy \in \R} 
    \cvar_{\alpha}\Big(
    \entr_\lambda
    \big(
        a \Delta S_1
        ~ \big| ~
        \Delta S_1 \sim \calN(\mu, \sigma)
    \big)
    ~ \Big| ~
    \mu \sim \calN(\hat{\mu}, \Sigma/N)
    \Big),
\end{align*}
is uniquely attained by
\[
\bfarr
=\frac{1}{\lambda}\,
\Bigl(1-\frac{\kappa_\alpha}{\sqrt{N}\,\|\Sigma^{-1/2}\hat\mu\|}\Bigr)_+\,
\Sigma^{-1}\,\hat\mu,
\]
where  \(\displaystyle \kappa_\alpha={\varphi(\Phi^{-1}(\alpha))}/{\alpha}\). 
Note that there is a ``zero-investment region''
\(\;\bfarr=0\) whenever
\(\|\Sigma^{-1/2}\hat\mu\|\le \kappa_\alpha/\sqrt{N}\),
and otherwise \(\bfarr\) points along \(\Sigma^{-1}\hat\mu\) with shrinkage factor \(\bigl(1-\kappa_\alpha/(\sqrt{N}\|\Sigma^{-1/2}\hat\mu\|)\bigr)/\lambda\).
\end{restatable}

\begin{figure}
    \centering
    \includegraphics[width=\textwidth]{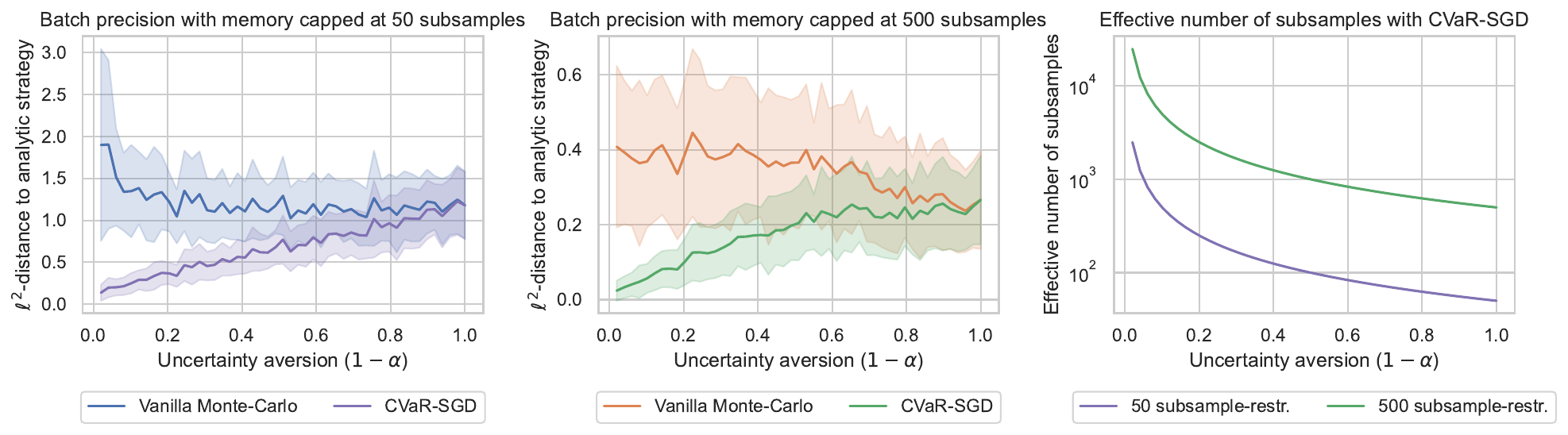}
    \caption{
    \oldchange{
    Approximation of the uncertainty-aware strategy in \Cref{lem:MDCVARStrategy}, with $\dest, \Sigma$ estimated based on the last $30$ daily returns of the $d=503$ components of the S\&P500 index. For the analytic strategy approximation, we compare two cases: the left panel uses $m = 50$ subsamples (each of size $n = 30$), while the middle panel increases this to $m = 500$. The improvement in precision achieved by the $\cvar$-SGD method (described in \Cref{sec:CVaRGradientDescent}) is due to its ability to effectively handle a larger number of subsamples for higher values of $\alpha$, see the figure on the right-hand side.
    }
    }
    \label{fig:ApproxHD}
\end{figure}


\paragraph{Approximation.} In our experiment, we first extract daily log‐returns for each of the $d=503$ stocks in the S\&P 500 over the last 30 trading days before 20th May 2025, and use these to compute the empirical mean $\hat\mu$ and covariance $\Sigma$.  As a benchmark, we use the analytic CVaR strategy of Lemma \ref{lem:MDCVARStrategy} applied to the estimates on these historical returns on a grid of values for $\alpha$. 

We then proceed to compute the numerical strategies, following the sub-sampling approach. First, we use $m=50$ (and $m=500$) subsamples of size $n=30$ combined with a standard gradient descent to approximate the uncertainty-aware strategy (see \Cref{fig:ApproxHD}). We then record the average Euclidean distance between the benchmark and each approximation as our error metric. We can see that the precision is significantly higher for larger numbers of subsamples $m$ and that the performance generally decreases with increasing $\alpha$. This is expected, as we know that $\cvar$ does not use an $\alpha$-fraction of the simulations. 

In both cases, we can reverse this trend using the $\cvar$-SGD, whilst maintaining the respective memory constraint. The precision remains largely the same for $\alpha$ close to 0, but improves significantly with increasing $\alpha$. This is due to the fact, that $\cvar$-SGD allows the usage of roughly $m / (1-\alpha)$ subsamples, requiring the same memory as the standard gradient descent with $m$ subsamples. This is visualized as the ``effective number of subsamples'' on the right hand side of \Cref{fig:ApproxHD}.

\subsection{Deep Learning Example - Deep Hedging on cliquet options}\label{sec:DH}
\begin{figure}
    \centering
    \includegraphics[width=\textwidth]{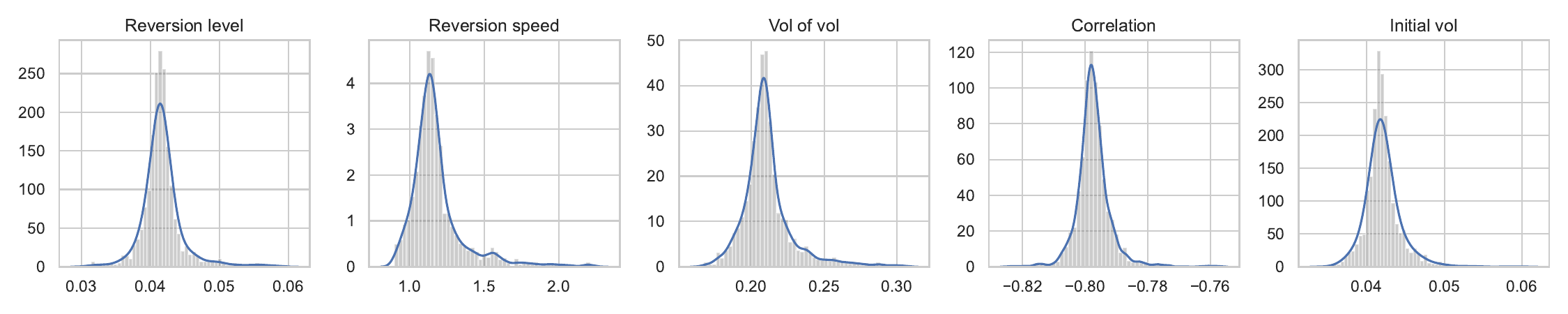}
    \caption{Test distributions for true underlying parameters, given the estimation of parameters in \Cref{tab:plug-in_parameters}. The distribution is motivated by 1-day changes in calibrated Heston models to the S\&P500 index option chain.}
    \label{fig:HestonPars}
\end{figure}

\begin{figure}[ht!]
    \centering
    \includegraphics[height=5cm]{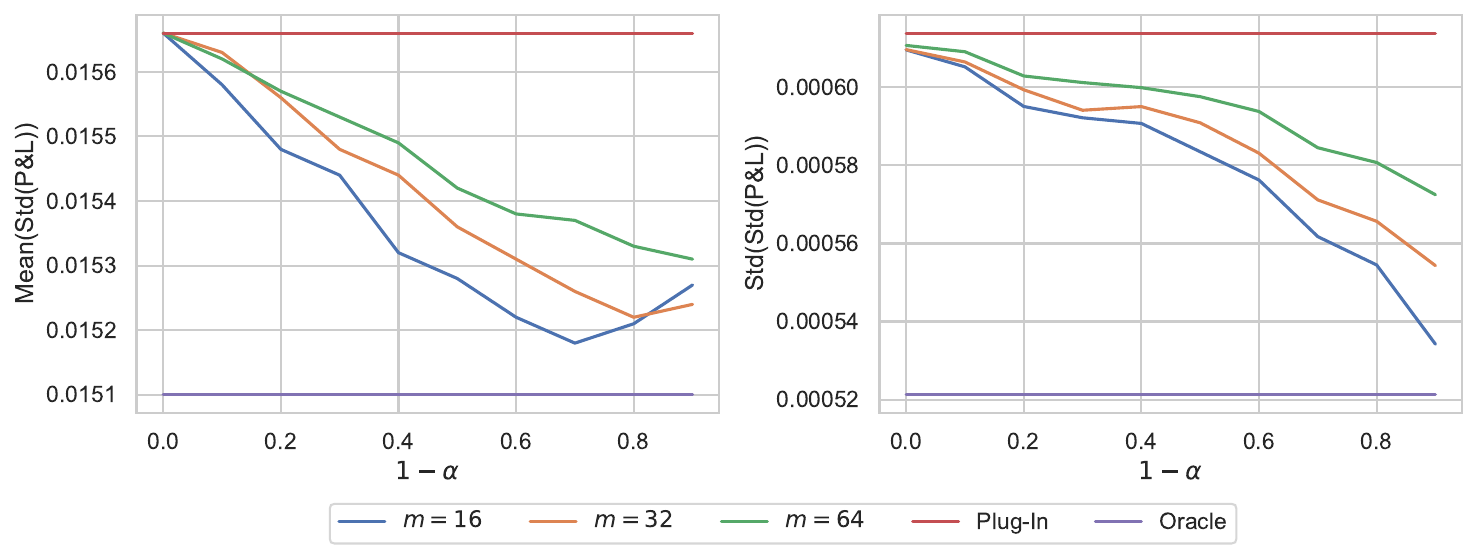}
    \caption{
    For every Heston parameter set, one can compute the objective, which is the standard deviation of the P\&L under this Heston model.
    Doing this for every Heston parameter set sampled from the test distribution $\smash{\tilde{\frakP}}$ (\Cref{fig:HestonPars}), one can compare the mean objective (left panel) and standard deviation of objective (right panel).
    The plot shows results from the plug-in, uncertainty-aware (for various subsample sizes $m \in \{16,32,64\}$), and oracle hedge. 
    Note, since the objective is chosen to be the standard deviation of P\&L, lower means (left panel) and lower standard deviations (right panel) are desired across the test distribution.
    It is apparent that uncertainty-aware hedges significantly increase the performance in both better average and less variability of the objective.
    }
    \label{fig:HedgeExampleSimulated}
\end{figure}


In this subsection, we demonstrate numerically that subsampling can be effectively employed to increase robustness in path-dependent tasks. To this end, we consider a hedging task for a cliquet option over a period of 120 trading days with a reset period of 40 days. The cliquet option payoff is defined as
\[
C_\mathrm{cliquet} = \sum_{i=1}^{3}  \max\left({S_{40i}} - {S_{40(i-1)}}, 0\right),
\]
where \(S_{40i}\) denotes the asset price at the \(i\)-th reset time.

The objective for this hedging task is to minimize the standard deviation of the profit-and-loss distribution. As the estimated model $\hat{\bbP}$, we select a Heston model with the plug-in parameter estimations shown in \Cref{tab:plug-in_parameters}. In our setup, the hedging strategy can only access the spot price process, along with the spot price at the most recent reset (which serves as the current strike price).

\begin{table}
\centering
\begin{tabular}{cccccc}
\toprule
Reversion speed          & 
Reversion level          & 
Volatility of volatility & 
Correlation              & 
Drift                    & 
Initial volatility       \\ 
\midrule
1.0   &
0.03  &
0.2   &
-0.8  &
0     &
0.03  \\
\bottomrule
\end{tabular}
\vspace{.2em}
\caption{Plug-in Heston model parameters for \Cref{sec:DH}.}
\label{tab:plug-in_parameters}
\end{table}

\oldchange{
\paragraph{Vanilla deep hedge.}
We first train a deep hedge using the plug-in Heston parameters.  The deep hedging framework \cite{Buehler2019}, replaces classical closed‐form hedging formulas with a data‐driven approach based on deep learning. While we refer to \cite{Buehler2019} for details, we repeat the core idea here. Rather than deriving a replicating strategy under a specific model, deep hedging parameterizes the hedging policy as a neural network
\[
\pi_\phi\;:\;\text{information at time $t$}\mapsto\text{asset position at time }t.
\]
We will refer to the information process---containing the current spot price and the most recent reset price---by \((h_t)_{t=0}^{120}\subseteq\R^2\). If $\Phi$ is the set of all parameters, i.e. $\phi \in \Phi$, we have $\StrategySpace = \mySet{\pi_\phi}{\phi \in \Phi}$.

Training proceeds by simulating a large number of market scenarios \(S=(S_t)_{t=0}^{120}\) from the estimated Heston model represented by $\hat{\bbP}$.  The policy \(\pi_\phi\) with parameters \(\phi \in \Phi\) is optimized by minimizing the standard deviation of the hedging profit-and-loss as follows
\[
\min_{\phi \in \Phi}\; \mathrm{Std}_{S\sim \hat \bbP} \Big[-C_\mathrm{cliquet}  + \sum_{t=1}^T \pi_\phi(h_t)\,\Delta S_t\Big].
\]
}

\oldchange{
\paragraph{Uncertainty-aware deep hedge using subsampling.}
Subsequently, we develop uncertainty-aware versions of the strategy by employing \(\cvar\) as the uncertainty measure with subsampling. Following the definition \eqref{id:RobustStrategy}, we see that we can use the deep hedging procedure with the modified objective
\[
\min_{\phi \in \Phi}\; \cvar_\alpha\left[
\widehat{\mathrm{Std}} \Big[
\Big(
    -C_\mathrm{cliquet}  + \sum_{t=1}^T \pi_\phi(h_t)\,\Delta S^{(i)}_t
\Big)_{i=1}^m
\Big]
\middle|
(S^{(i)})_{i=1}^m \sim \hat\bbP^m
\right] 
\]
to obtain the uncertainty-aware strategy, where $\widehat{\mathrm{Std}}$ estimates the standard deviation empirically on the $m$-subsample.

In our experiment, we train these uncertainty-aware strategies using various uncertainty aversion parameters \(\alpha \in [0,1]\) and subsample sizes \(m \in \{16,\,32,\,64\}\).
}

\paragraph{Experiment results.} To assess the out-of-sample performance of the different strategies\oldchange{, we intend to monitor the hedges performance for the true underlying dynamics. Since we chose a set of fixed parameters (see \Cref{tab:plug-in_parameters}), we can test the model now under a variety of parameters which \textit{could} be the true underlying model.} For this, we specify a distribution of Heston models $\tilde{\frakP}$---each weighted by its likelihood of representing the "true" model. This distribution is calibrated to realistically reflect typical one-day Heston parameter changes of the S\&P500 index and is illustrated in \Cref{fig:HestonPars}. 

The results of our analysis, which compare the vanilla deep hedge against the uncertainty-aware strategies, are presented in \Cref{fig:HedgeExampleSimulated}. 
\oldchange{
We can see that the robusitified hedges consistently outperform their plug-in counterparts in both mean and variance of the resulting objectives on the test distribution $\tilde\frakP$.
When comparing the subsample sizes $m$ displayed, we notice that $m=16$ seems the most effective, indicating the need for a rather substantial robustification. This is underlined by the fact that strategies with $m=32$ and $m=64$ consistenly improve for decreasing values of $\alpha$, rendering strong uncertainty-aversion preferences as the most effective.
}

To showcase that the uncertainty-aware strategy derives meaningful performance improvements, we include an ``Oracle'' strategy, which is the mixture approach on the test-distribution \Cref{fig:HestonPars}.
\oldchange{
This can be achieved by training a deep hedging strategy with the following objective
\[
\min_{\phi \in \Phi}\; \E_{\bbP \sim \tilde{\frakP}}\left[\mathrm{Std}_{S\sim \bbP} \Big[-C_\mathrm{cliquet}  + \sum_{t=1}^T \pi_\phi(h_t)\,\Delta S_t\Big]\right].
\]
}

Clearly, by using the test-distribution, this approach accesses information that is not available to the agent, therefore its out-performance is not surprising. We would like to highlight however, that a reasonably calibrated uncertainty-aware strategy based on subsampling attains comparable performance enhancements when comparing the mean objective (and standard deviation of the objective) on the test-distribution.

\bibliographystyle{plain}
\bibliography{references}  

\appendix
\newpage
\section{Basics on Risk Measures}

Consider the probability space $(\Omega, \calF, \bbP)$ and let $X \in L^0(\Omega, \calF, \bbP)$. Then, the entropy \cite{FoellmerSchied04} for risk aversion $\lambda>0$ is defined as follows
	\begin{align}\label{def:Entropy}
		\entr_\lambda(X) := - \frac{1}{\lambda} \log\left( \myE{\bbP}{\exp \left( - \lambda X \right)} \right).
	\end{align}

For normal random variables, it is useful to exchange the entropic risk-measure with the mean-variance operator \cite{de1940problema}, since it offers higher numerical stability.
	 The mean-variance for risk-aversion $\lambda>0$ is defined by 
	\begin{align}\label{def:MV}
		\mv_\lambda(X) := \myE{\bbP}{X} - \frac{1}{2}\lambda \myV{\bbP}{X}.
	\end{align}

If $X$ is normally distributed, the Laplace transform equals \begin{align}
    \myE{\bbP}{\exp \left( - \lambda X \right)}=\exp \left( - \lambda \myE{\bbP}{X} + \nicefrac{\lambda}2 \myV{\bbP}{X}\right), \label{id:MVisENT}
\end{align} 
such that mean-variance and negative entropy coincide.



	In the above setting, the conditional value at risk (CVaR) \cite{FoellmerSchied04} is defined for a risk-aversion parameter $\alpha \in [0,1]$ as 
	\begin{align}
            - \cvar_\alpha(X) := \myCond{\bbP}{X}{X \leq \mathrm{VaR}_\alpha(X)}, \label{id:CVaR}
	\end{align}
	where for $\gamma \in [0,1]$
	\begin{align*}
		\mathrm{VaR}_{\gamma}(X)= - \inf \left\lbrace x \in \mathbb{R} : F_{X}(x) > \gamma \right\rbrace.
	\end{align*}


\begin{rem}
	Note we can rewrite 
	$$
	\cvar(X) = \frac{1}{\alpha} \left(  \mathbb E[{X} 1_{\lbrace X \leq x_\alpha \rbrace}]+ x_\alpha(\alpha-\mathbb P(X<x_\alpha)) \right),
	$$
	where $x_\alpha$ is the lower $\alpha$-quantile. This particularly simplifies for continuous distribution functions, since the term $x_\alpha(\alpha-\mathbb P(X<x_\alpha)) = 0$ vanishes.
\end{rem}


\section{Results for \texorpdfstring{\Cref{sec:ComparingAnalyticallyIIDNormal}}{Section 3}} \label{app:Iid}

\begin{lem}\label{lem:SolveEntr}
    For $X \sim \calN(\nu, \tau^2)$ with $\tau, \nu, \lambda \in \R$, 
    \begin{align*}
        \argmax_{a \in \R}\entr_\lambda(aX) = \frac{\nu}{\lambda \tau^2}
    \end{align*}
\end{lem}
\begin{proof}
    As remarked above, under normality, the entropy coincides with mean-variance, such that, the left-hand side reduces to $a\nu - \lambda a^2\tau^2/2$ which is optimized by the right-hand side.
\end{proof}

\begin{lem}\label{lem:OOSPScaledPlug-inReduction}
Assume that $\Delta S_1 \sim \calN(\mu,\sigma^2)$ and $\hat \mu \sim \calN(\mu,\sigma^2N^{-1})$. Then,
\begin{align*}
    \entr_\lambda\left( x \dest \Delta \pS_{1}\right) 
    & =
    \entr_\lambda\left( x \mu \Delta \pS_{1}\right) 
    -
    \frac{\lambda}{2}
    \myV{\bbP}{x(\dest - \mu) \Delta \pS_{1}} \\
    &=
    x \mu^2
    \left(1-  \frac{\lambda}{2}x\sigma^2\right)
    -
    \frac{\lambda }{2N} x^2\sigma^2(\mu^2 + \sigma^2).
\end{align*}
\end{lem}

\begin{proof}
We start with the first equation:
due to equivalence of $\entr$ and $\mv$ for normal random variables,   
\begin{align*}
	\entr_\lambda\left( x \dest \Delta \pS_{1}\right)
	&= \mv_\lambda\left(  x \dest \Delta \pS_{1}\right) \\
        &= \myE{\bbP}{ x \dest \Delta \pS_1} - \frac{\lambda}{2} \myV{\bbP}{ x \dest \Delta \pS_1} \\
        &= x\myE{\bbP}{\mu \Delta \pS_{1}} 
        - \frac{x^2\lambda}{2} \myV{\bbP}{ \mu \Delta \pS_{1} +  (\dest - \mu) \Delta \pS_{1}}.
\end{align*}
Next, observe that
\begin{align}
        &\myV{\bbP}{ \mu \Delta \pS_{1} + (\dest - \mu) \Delta \pS_{1}} \notag \\
        &\quad = 
        \myE{\bbP}{ (\mu \Delta \pS_{1})^2 + 2 \mu  (\dest - \mu) \Delta \pS_{1}^2 + (\dest - \mu)^2 \Delta \pS_{1}^2}
        -
        \myE{\bbP}{ \mu \Delta \pS_{1} + (\dest - \mu) \Delta \pS_{1}}^2
        \notag \\
        &\quad = 
        \myE{\bbP}{ (\mu \Delta \pS_{1})^2}  + 
        \myE{\bbP}{(\dest - \mu)^2 \Delta \pS_{1}^2}
        -
        \myE{\bbP}{ \mu \Delta \pS_{1}}^2
        -
        \myE{\bbP}{(\dest - \mu) \Delta \pS_{1}}^2
        \notag \\
    &\quad = \myV{\bbP}{ \mu \Delta \pS_{1}} + 
    \myV{\bbP}{(\dest - \mu) \Delta \pS_{1}}. \label{id:OospPlug-inPlug-in_V}
\end{align}
The second equation follows by observing that
by iterated conditional expectations,
 \begin{align} 
         \mu \myE{\bbP}{  \Delta \pS_{1}} =  \myE{\bbP}{ \Delta \pS_1} = 
         \mu^2, 
         \label{id:OospPlug-inPlug-in_E}
 \end{align}
together with
    \begin{align*}
        \myV{\bbP}{(\dest - \mu) \Delta \pS_{1}} 
        = 
        \myE{\bbP}{(\dest - \mu)^2} 
        \myE{\bbP}{\Delta \pS_{1}^2} 
        = 
        \frac{\sigma^2}{N}
        (\mu^2 + \sigma^2).
    \end{align*}
\end{proof}

\begin{lem}
\label{lem:OospPlug-inEmperic}
    The ``out-of-sample'' performance of $\bfan$ as in \eqref{id:Plug-inStrategy}, under the assumptions of \Cref{lem:OOSPScaledPlug-inReduction} is
    \begin{align}
        \mv^\bbP_\lambda(\bfan\Delta\pS_{1}) = \frac{\mu^2}{2\lambda \sigma^2} - \frac{\mu^2 + \sigma^2}{2 N \lambda \sigma^2}         \label{id:AnalyticalOOSPPlug-inMV}
    \end{align}
\end{lem}

\begin{proof}
The explicit form of the strategy is given by \Cref{lem:SolveEntr} and the statement follows from \Cref{lem:OOSPScaledPlug-inReduction} with $x = 1/ (\lambda \sigma^2 )$.
\end{proof}

\begin{lem}\label{lem:OOSPJoint}
    The ``out-of-sample'' performance of $\bfac$  as in \eqref{id:MixtureIid}, under the assumptions of \Cref{lem:OOSPScaledPlug-inReduction}, is 
    \begin{align*}
        \entr_\lambda(\bfac\Delta\pS_{1}) =
        \frac{\mu^2N}{\lambda\sigma^2 (N + 1)} 
        \left(1-\frac{\lambda}{2}\sigma^2\right)
        -
        \frac{N(\mu^2 + \sigma^2)}{2\lambda\sigma^2 (N + 1)^2}
    \end{align*}
\end{lem}
\begin{proof}
    This follows directly from \Cref{lem:SolveEntr} and \Cref{lem:OOSPScaledPlug-inReduction}.
\end{proof}

\OospUAIidEntrEntr*

\begin{proof}
By the same arguments as in \Cref{lem:OospUAIidEntrEntr} and $\bfa \in \calA$,
\begin{align*}
\mv^\frakP_{\lambda'} \left( \hat{\mv}_\lambda \left( \lbrace \bfa (\Delta S_t + \peps) \rbrace_{t \in \timesteps} \right) \right)
    &= \mv^\frakP_{\lambda'}\left( \bfa( \dest + \peps) - \frac{\lambda}{2}\bfa^2 \sest \right)
\end{align*}
and thus
\begin{align*}
    \mv^\frakP_{\lambda'}\left( \bfa( \dest + \peps) - \frac{\lambda}{2}\bfa^2 \sest \right) &= \myE{\frakP}{\bfa( \dest + \peps) - \frac{\lambda}{2} \bfa^2 \sest } - \frac{\lambda'}{2} \myV{\frakP}{\bfa( \dest + \peps) - \frac{\lambda}{2} \bfa^2 \sest} \\
    &= \bfa \dest - \frac{\lambda}{2} \bfa^2 \sest  - \frac{\lambda'}{2} \bfa^2\myV{\frakP}{ \peps} \\
    &= \bfa \dest - \frac{\lambda}{2} \bfa^2 \sest \left( 1 + \frac{\lambda'}{\lambda N}\right).
\end{align*}
Hence, the strategy $\bfar$ maximizing the objective can be expressed as
\begin{align*}
    \bfar :=  \frac{\dest}{\lambda  \sest(1 + \frac{\lambda'}{\lambda N})}.
\end{align*}
A straightforward computation shows that
\begin{align*}
    \bfar = \frac{\lambda N}{\lambda N + \lambda'} \bfan.
\end{align*}

By employing results \eqref{id:OospPlug-inPlug-in_E} and \eqref{id:OospPlug-inPlug-in_V}, we receive for $x \in \R$
\begin{align*}
	\mv^\bbP_\lambda\left(x \, \bfan \Delta \pS_{1}\right)
        &=\mv^\bbP_\lambda\left(x \frac{\dest}{\lambda \sigma^2} \Delta \pS_{1}\right)\\
        &=x \myE{\bbP}{  \frac{\dest}{\lambda \sigma^2} \Delta \pS_{1}} - \frac{\lambda}{2} x^2 \myV{\bbP}{ \frac{\dest}{\lambda \sigma^2} \Delta \pS_{1}} \\
        &=x \myE{\bbP}{ \frac{\mu}{\lambda \sigma^2} \Delta \pS_{1}}
        - \frac{\lambda}{2} x^2\left(\myV{\bbP}{ \frac{\mu}{\lambda \sigma^2} \Delta \pS_{1}} +   \frac{\mu^2 + \sigma^2}{N \lambda^2 \sigma^2}\right) \\
        &=x \frac{\mu^2}{\lambda \sigma^2}
        - x^2 \frac{2N\mu^2 + \mu^2 + \sigma^2}{2N \lambda \sigma^2} \\
        &= \frac{x}{\lambda \sigma^2} \left(\mu^2 \left( 1 - \frac{x}{2} \right) - \frac{x(\mu^2 + \sigma^2)}{2N} \right)
\end{align*}
and hence for $x_N := \lambda N / (\lambda N + \lambda')$,
\begin{align*}
    \mv^\bbP_\lambda\left( \bfar \Delta \pS_{1}\right) 
    &= \frac{x_N}{\lambda \sigma^2} \left(\mu^2 \left( 1 - \frac{x_N}{2} \right) - \frac{x_N(\mu^2 + \sigma^2)}{2N} \right).
\end{align*}
\end{proof}

\begin{lem}
\label{lem:UaIsBetter}
    The uncertainty-aware out-performs the plug-in policy if and only if 
    \begin{align*}
         \mu^2 \frac{(N-1)\lambda'  - 2N \lambda }{2N\lambda  + \lambda'} < \sigma^2, 
    \end{align*}
    implying that the robustification is necessarily successful, whenever 
    \begin{align*}
        \lambda' < \frac{2N}{N-1}\lambda.
    \end{align*}
\end{lem}
\begin{proof}
Set 
\[
x_N = \frac{\lambda N}{\lambda N + \lambda'}.
\]
The out‐of‐sample value of the uncertainty‐aware policy is
\[
\frac{x_N}{\lambda\sigma^2}\Bigl[\mu^2\Bigl(1-\tfrac{x_N}{2}\Bigr) - \frac{x_N(\mu^2+\sigma^2)}{2N}\Bigr],
\]
while the plug‐in policy yields
\[
\frac{\mu^2}{2\lambda\sigma^2} - \frac{\mu^2+\sigma^2}{2N\lambda\sigma^2}.
\]
Thus the former exceeds the latter exactly when
\[
\frac{x_N}{\lambda\sigma^2}\Bigl[\mu^2(1-\tfrac{x_N}{2}) - \tfrac{x_N(\mu^2+\sigma^2)}{2N}\Bigr]
>
\frac{\mu^2}{2\lambda\sigma^2} - \frac{\mu^2+\sigma^2}{2N\lambda\sigma^2}.
\]
Multiply both sides by $2N\lambda\sigma^2>0$:
\[
2N x_N\mu^2\Bigl(1-\tfrac{x_N}{2}\Bigr)
- x_N^2(\mu^2+\sigma^2)
>
N\mu^2 - (\mu^2+\sigma^2).
\]
Expand and collect terms:
\[
2N x_N\mu^2 - N x_N^2\mu^2 - x_N^2\mu^2 - x_N^2\sigma^2
>
N\mu^2 - \mu^2 - \sigma^2,
\]
\[
\mu^2\bigl(2N x_N - N x_N^2 - x_N^2 - N + 1\bigr)
>
-\,\sigma^2(1 - x_N^2).
\]
Note that
\[
2N x_N - N x_N^2 - x_N^2 - N + 1
= -\,N(x_N^2 - 2x_N + 1) - (x_N^2 - 1)
= -\,(1 - x_N)\bigl(N(1 - x_N) - (1 + x_N)\bigr).
\]
Hence
\[
-\,\mu^2(1 - x_N)\bigl(N(1 - x_N) - (1 + x_N)\bigr)
>
-\,\sigma^2(1 - x_N^2),
\]
\[
\mu^2(1 - x_N)\bigl(N(1 - x_N) - (1 + x_N)\bigr)
<
\sigma^2(1 - x_N^2).
\]
Since $0 < x_N < 1$, divide by $(1 - x_N)(1 + x_N)$ to get
\[
\mu^2\frac{N(1 - x_N) - (1 + x_N)}{\,1 + x_N\,} < \sigma^2.
\]
Finally, substitute 
\[
1 - x_N = \frac{\lambda'}{\lambda N + \lambda'}, 
\quad
1 + x_N = \frac{2\lambda N + \lambda'}{\lambda N + \lambda'}.
\]
Then
\[
N(1 - x_N) - (1 + x_N)
= \frac{N\lambda' - (2\lambda N + \lambda')}{\lambda N + \lambda'}
= \frac{(N-1)\lambda' - 2N\lambda}{\lambda N + \lambda'},
\quad
1 + x_N = \frac{2\lambda N + \lambda'}{\lambda N + \lambda'}.
\]
Thus
\[
\frac{N(1 - x_N) - (1 + x_N)}{\,1 + x_N\,}
= \frac{(N-1)\lambda' - 2N\lambda}{2N\lambda + \lambda'},
\]
and the inequality becomes
\[
\mu^2 \frac{(N-1)\lambda' - 2N\lambda}{2N\lambda + \lambda'} < \sigma^2,
\]
as claimed. Finally, since $2N\lambda + \lambda'>0$, this holds in particular whenever $(N-1)\lambda' - 2N\lambda < 0$, i.e.\ $$\lambda' < \frac{2N}{N-1}\lambda.$$
\end{proof}

\OospUAIidEntrCVaR*

\begin{proof}
Since we have chosen $\peps \sim \calN(0, \sest/N)$, we can compute for every $\bfa \in \calA$
\begin{align*}
    \cvar^\frakP_{\alpha} \left( \hat{\mv}_\lambda \left( \lbrace \bfa (\Delta S_t + \peps) \rbrace_{t \in \timesteps} \right) \right) &= \cvar^\frakP_{\alpha}\left( \bfa (\dest + \peps) - \frac{\lambda}{2}\bfa^2\sest \right) \\
    &= \left( \bfa \dest - \frac{\lambda}{2}\bfa^2 \sest \right) - A \vert \bfa \vert \sqrt{\frac{\sest}{N}} 
\end{align*}
where $A := \phi(\Phi^{-1}(\alpha))/\alpha$. Here, $\phi(\cdot)$ denotes the probability density function and $\Phi^{-1}(\cdot)$ the quantile function of the standard normal distribution. Note that we have to use the absolute value of $\bfa$ to apply that formula, since negative factorized random variables have their worst scenarios at the other tail. 

Point-wise maximization of the above expression results in 
\begin{align}
    \bfarr = \sign(\dest)\frac{\Big(\vert \dest \vert -   A \sqrt{\frac{\sest}{N}}\Big)_+}{\lambda \sest}
\end{align}
where $\sign(\cdot)$ is the sign operator and $(\cdot)_+$ is the restriction to the positive orthant.

Eventually, we calculate again the out-of-sample performance
\begin{align}
	\mv^\bbP_\lambda\left( \bfarr \Delta \pS_{1}\right) &=
	\mv^\bbP_\lambda\left(\sign(\dest)\frac{\Big(\vert \dest \vert -   A \sqrt{\frac{\sest}{N}}\Big)_+}{\lambda \sest} \Delta \pS_{1}\right). \label{id:OOSPcvar}
\end{align}
Using the assumption $\sest \equiv \sigma^2$, we see that
\begin{align*}
    & \mv^\bbP_\lambda\left(\sign(\dest)\frac{\Big(\vert \dest \vert -   A \sqrt{\frac{\sest}{N}}\Big)_+}{\lambda \sest} \Delta \pS_{1}\right) \\
     = & \mu \myE{\bbP}{\sign(\dest)\frac{\Big(\vert \dest \vert - A \sqrt{\frac{\sest}{N}}\Big)_+}{\lambda \sest}} 
    - \frac{\lambda \sigma^2}{2}\myE{\bbP}{\left(\frac{\Big(\vert \dest \vert - A \sqrt{\frac{\sest}{N}}\Big)_+}{\lambda \sest}\right)^2} 
    - \frac{\lambda \mu^2}{2}\myE{\bbP}{\sign(\dest)\frac{\Big(\vert \dest \vert - A \sqrt{\frac{\sest}{N}}\Big)_+}{\lambda \sest}}
\end{align*}

We compute 
\begin{align*}
    &2\myE{\bbP}{\sign(\dest)\Big(\vert \dest \vert - A \sqrt{\frac{\sest}{N}}\Big)_+}\\
    &=\frac{\mu ^2 \sqrt{N} \erf\left(\frac{\sqrt{\left(A \sigma +\mu  \sqrt{N}\right)^2}}{\sqrt{2} \sigma }\right)}{\sqrt{\left(A \sigma +\mu  \sqrt{N}\right)^2}}
    +\mu  \erf\left(\frac{\mu  \sqrt{N}-A \sigma }{\sqrt{2} \sigma }\right)
    - \frac{A \mu  \sqrt{N} \sigma  \erf\left(\frac{\sqrt{\left(A \sigma +\mu  \sqrt{N}\right)^2}}{\sqrt{2} \sigma }\right)}{\sqrt{N \left(A \sigma +\mu  \sqrt{N}\right)^2}} \\
    &-\frac{A \sigma  \erf\left(\frac{\mu  \sqrt{N}-A \sigma }{\sqrt{2} \sigma }\right)}{\sqrt{N}}
    +\frac{A \sigma  \erfc\left(\frac{A \sigma +\mu  \sqrt{N}}{\sqrt{2} \sigma }\right)}{\sqrt{N}}
    - \frac{\sqrt{\frac{2}{\pi }} \sigma  e^{-\frac{\left(A \sigma +\mu  \sqrt{N}\right)^2}{2 \sigma ^2}}}{\sqrt{N}}
    + \frac{\sqrt{\frac{2}{\pi }} \sigma  e^{-\frac{\left(\mu  \sqrt{N}-A \sigma \right)^2}{2 \sigma ^2}}}{\sqrt{N}}
    - \frac{A \sigma }{\sqrt{N}}\\
    &+\frac{\sqrt{\mu ^2 N} \erf\left(\frac{\sqrt{\mu ^2 N}}{\sqrt{2} \sigma }\right)}{\sqrt{N}}
    - \mu  \erf\left(\frac{\mu  \sqrt{N}}{\sqrt{2} \sigma }\right)
    + 2 \mu \Bigg)
\end{align*}
where $\erf(\cdot) := 2/\sqrt{\pi} \int_0^\cdot \exp(-x^2) dx$  and $\erfc(\cdot) = 1 - \erf(\cdot)$. Furthermore,
\begin{align*}
    2N\myE{\bbP}{\left(\Big(\vert \dest \vert - A \sqrt{\frac{\sest}{N}}\Big)_+\right)^2} &= \left(\left(A^2+1\right) \sigma ^2-2 A \mu  \sqrt{N} \sigma +\mu ^2 N\right) \erfc\left(\frac{A \sigma -\mu  \sqrt{N}}{\sqrt{2} \sigma }\right) \\
    & + \left(\left(A^2+1\right) \sigma ^2+2 A \mu  \sqrt{N} \sigma +\mu ^2 N\right) \erfc\left(\frac{A \sigma +\mu  \sqrt{N}}{\sqrt{2} \sigma }\right) \\
    & - \sqrt{\frac{2}{\pi }} \sigma  e^{-\frac{\left(A \sigma +\mu  \sqrt{N}\right)^2}{2 \sigma ^2}} \left(A \sigma  \left(e^{\frac{2 A \mu  \sqrt{N}}{\sigma }}+1\right)-\mu  \sqrt{N} \left(e^{\frac{2 A \mu  \sqrt{N}}{\sigma }}-1\right)\right).
\end{align*}
Thus, we obtain a closed form expression of \eqref{id:OOSPcvar}.
\end{proof}

\UAStratLimits*
\begin{proof}[Proof of Corollary~\ref{cor:UAStratLimits}]
We begin with the entropic uncertainty-aware strategy from Lemma~\ref{lem:OospUAIidEntrEntr}, which has the analytic form
\[
\bfar = \frac{\lambda N}{\lambda N + \lambda'} \bfan =: x_N \bfan,
\]
where $x_N := \frac{\lambda N}{\lambda N + \lambda'}$.

\begin{itemize}
    \item As $\lambda' \to 0$, we have $x_N \to 1$, hence $\bfar \to \bfan$.
    \item As $\lambda' \to \infty$, we have $x_N \to 0$, hence $\bfar \to 0$.
    \item As $N \to \infty$, we have $x_N \to 1$, so again $\bfar \to \bfan$.
    
    But since $\bfan = \dest  / (\lambda \sest)$ and $\dest \to \mu$, $\sest \to \sigma^2$ as $N \to \infty$, we conclude $\bfan \to \bfao := \mu / (\lambda \sigma^2)$.
    Therefore, $\bfar \to \bfao$ as $N \to \infty$.
\end{itemize}

Next, for the CVaR-based uncertainty-aware strategy from Lemma~\ref{lem:OospUAIidEntrCVaR}, we have
\[
\bfarr = \left( \bfan - \sign(\dest)\frac{A \sqrt{\frac{\sest}{N}}}{\lambda \sest} \right)\myind{\dest^2 \geq A^2 \frac{\sest}{N}},
\]
where $A := \phi(\Phi^{-1}(\alpha)) / \alpha$.

\begin{itemize}
    \item As $\alpha \to 1$, we have $A \to 0$, so the indicator condition $\dest^2 \geq A^2 \sest / N$ is always satisfied and the correction term vanishes. Therefore, $\bfarr \to \bfan$.
    \item As $\alpha \to 0$, we have $A \to \infty$, so the indicator becomes zero with probability one, hence $\bfarr \equiv 0$.
    \item As $N \to \infty$, the term $A \sqrt{\sest / N} \to 0$ and $\dest \to \mu$, $\sest \to \sigma^2$, thus the indicator becomes 1 for large $N$ and the correction vanishes. Therefore, $\bfarr \to \bfao := \mu / (\lambda \sigma^2)$.
\end{itemize}
\end{proof}

\section{Results for Section \texorpdfstring{\Cref{sec:CVaRGradientDescent}}{Section 5}}
\subsection{Details of \texorpdfstring{\Cref{thm:CVARSGD-randomk}}{Theorem 5.2}}\label{app:ProofCVaRSG}

\CVARSDGTHM*

\begin{proof}
\textbf{Convexity and Lipschitz continuity.} Using the Rockafellar--Uryasev representation of $\cvar$ \cite{RockafellarUryasev00},
\[
\cvar_\alpha(Z) = \min_{\zeta \in \mathbb{R}} \left\{ \zeta + \frac{1}{\alpha} \mathbb{E}\left[ (Z - \zeta)_+ \right] \right\},
\]
we write
\[
\mathcal{L}(\phi) = \min_{\zeta \in \mathbb{R}} \left\{ \zeta + \frac{1}{\alpha} \mathbb{E}_{\bbP \sim \ModelDist} \left[ (J(X(a_\phi), \bbP) - \zeta)_+ \right] \right\}.
\]
Because $J(X(a_\phi), \bbP)$ is convex in $\phi$ and the hinge function $(\cdot - \zeta)_+$ is jointly convex, the full objective is convex in $\phi$. The bound on $\|\nabla_\phi J\|$ implies that $\mathcal{L}$ is $\rho$-Lipschitz.

\textbf{Unbiased sub-gradient via random $k$.} Let $J_\phi(\bbP) := J(X(a_\phi), \bbP)$ and let $\zeta_\phi := \mathrm{VaR}_\alpha(J_\phi)$. Then,
\[
\nabla_\phi \mathcal{L}(\phi) = \frac{1}{\alpha} \mathbb{E}_{\bbP \sim \ModelDist} \left[ \nabla_\phi J(X(a_\phi), \bbP) \cdot \mathbf{1}_{J_\phi(\bbP) \leq \zeta_\phi} \right].
\]
We now draw $m$ i.i.d. samples $\bbP_1, \dots, \bbP_m \sim \ModelDist$, compute $J_i := J(X(a_\phi), \bbP_i)$ and $g_i := \nabla_\phi J(X(a_\phi), \bbP_i)$, and recall that $k$ is a random integer with $\mathbb{E}[k] = \alpha m$.
Let $I(\phi)$ be the indices of the $k$ smallest $J_i$. Define the stochastic gradient estimator
\[
\widehat{g}(\phi) := \frac{1}{k} \sum_{i \in I(\phi)} g_i.
\]
By symmetry and independence, the distribution of the $k$ smallest elements ensures
\[
\mathbb{E}[\widehat{g}(\phi)] = \frac{1}{\alpha} \mathbb{E}_{\bbP \sim \ModelDist} \left[ \nabla_\phi J(X(a_\phi), \bbP) \cdot \mathbf{1}_{J_\phi(\bbP) \leq \zeta_\phi} \right] = \nabla_\phi \mathcal{L}(\phi).
\]
Hence, $\widehat{g}(\phi)$ is an unbiased stochastic sub-gradient. Its norm is bounded by $\rho$ by assumption.

\textbf{Apply SGD convergence.} All conditions of \cite[Theorem 14.8]{ShalevBen14} are satisfied: convex domain, unbiased sub-gradient with norm $\leq \rho$, Lipschitz objective, and learning rate $\eta_t = {B}/({\rho \sqrt{t}})$. The theorem then yields the expected bound
\[
\mathbb{E}[\mathcal{L}(\bar{\phi}_t)] - \min_{\phi \in \Phi} \mathcal{L}(\phi) \leq \frac{B \rho}{\sqrt{t}}.
\]
\end{proof}

\subsection{Results for \texorpdfstring{\Cref{sec:HighDim}}{Section 5.1}}\label{app:ProofMutliDimensional}
We compute the objective of  example of \Cref{sec:HighDim}
\begin{lem}\label{lem:HighDimObjective}
Assume that $\Delta S_t \;\sim\;\mathcal N_d(\hat \mu,\Sigma)$. %
Then, the  entropy  is given by
\begin{align}\label{temp:1900}
    J(X(a))\;=\;a^\top\hat\mu \;-\;\frac{\lambda}{2}\,a^\top\Sigma\,a.
\end{align}
\end{lem}
\begin{proof}
Since $X(a)$ is normally distributed, the entropy coincides with the mean-variance objective, as already remarked.
Note that $X(a)=a^\top \Delta S_t\sim \calN(a^\top\hat\mu,a^\top\Sigma\,a).$ The result then follows by Equation \eqref{def:MV}.
\end{proof}

\begin{lem}
\label{lem:plug-in_hd}
The plug-in portfolio as defined in \Cref{id:Plug-inStrategy} for the above setting is
\[
\bfan
=\frac{1}{\lambda}\,\Sigma^{-1}\, \dest.
\]
\end{lem}
\begin{proof}
    This result is obtained by maximizing Equation \eqref{temp:1900} in $a$. This equation is quadratic in $a$, such that the maximum is given by the first order condition
    $$ \hat \mu - \lambda  \Sigma a= 0.$$
\end{proof}

\MDCVARStrategy*
\begin{proof}
We compute
    \begin{align*}
    \bfarr 
    & = \argmax_{\Strategy \in \R} 
    \cvar_{\alpha}\Big(
    \entr_\lambda
    \big(
        a \Delta S_1
        ~ \big| ~
        S_1 \sim \calN(\mu, \sigma)
    \big)
    ~ \Big| ~
    \mu \sim \calN(\hat{\mu}, \Sigma/N)
    \Big)\\
    & = \argmax_{\Strategy \in \R} 
    \cvar_{\alpha}\Big(
    a^\intercal\mu \;-\;\frac{\lambda}{2}\,a^\intercal\Sigma\,a
    ~ \Big| ~
    \mu \sim \calN(\hat{\mu}, \Sigma/N)
    \Big)\\
    & = \argmax_{\Strategy \in \R} 
    \cvar_{\alpha}\Big(
    a^\intercal\mu 
    ~ \Big| ~
    \mu \sim \calN(\hat{\mu}, \Sigma/N)
    \Big)\;-\;\frac{\lambda}{2}\,a^\intercal\Sigma\,a\\
    & = \argmax_{\Strategy \in \R} 
    \underbrace{
    a^\top\hat\mu
    -\kappa_\alpha\sqrt{\tfrac{a^\top\Sigma a}{N}}
    -\frac{\lambda}{2}\,a^\top\Sigma\,a}_{:= \widetilde J(a)},
\end{align*}
where the last line follows from $\cvar_\alpha$ on normal random variables.

\textbf{(i) Zero‐investment region at \(a=0\).}  The subgradient of \(\widetilde J\) at \(0\) is
\[
\partial\widetilde J(0)
=\Bigl\{\hat\mu-\kappa_\alpha\tfrac{\Sigma^{1/2}u}{\sqrt N}\;\Big|\;\|u\|=1\Bigr\}.
\]
Hence \(0\) is optimal iff 
\(\max_{\|u\|=1}\langle\hat\mu,u\rangle\le \kappa_\alpha/\sqrt N\max_{\|u\|=1}\|\Sigma^{1/2}u\|\),
i.e.\ \(\|\Sigma^{-1/2}\hat\mu\|\le \kappa_\alpha/\sqrt N\).

\textbf{(ii) Outside the zero-investment region.}  For \(\|\Sigma^{-1/2}\hat\mu\|>\kappa_\alpha/\sqrt N\), symmetry implies the maximizer lies on the ray
\(\;a=c\,\Sigma^{-1}\hat\mu,\;c>0.\)
Set \(m=\|\Sigma^{-1/2}\hat\mu\|\), so
\[
a^\top\hat\mu=c\,m^2,
\qquad
\sqrt{a^\top\Sigma a}=c\,m.
\]
Then
\[
\widetilde J(c)
= c\,m^2
-\kappa_\alpha\frac{c\,m}{\sqrt N}
-\frac{\lambda}{2}\,c^2\,m^2.
\]
Differentiating w.r.t.\ \(c\) and setting to zero yields
\[
m^2-\kappa_\alpha\frac{m}{\sqrt N}-\lambda\,c\,m^2=0
\quad\Longrightarrow\quad
c^*
=\frac{1}{\lambda}\Bigl(1-\frac{\kappa_\alpha}{\sqrt{N}\,m}\Bigr).
\]
Since \(c^*>0\) precisely when \(m>\kappa_\alpha/\sqrt N\), we conclude
\[
\bfarr
=c^*\,\Sigma^{-1}\hat\mu
=\frac{1}{\lambda}
\Bigl(1-\frac{\kappa_\alpha}{\sqrt{N}\,\|\Sigma^{-1/2}\hat\mu\|}\Bigr)_+
\Sigma^{-1}\hat\mu,
\]
as claimed.
\end{proof}





\end{document}